\documentclass[fleqn,usenatbib]{mnras}
\usepackage{newtxtext,newtxmath}
\usepackage{url}
\usepackage{lmodern}        
\usepackage[normalem]{ulem}
\usepackage{textcomp}
\hypersetup{breaklinks=true}
\usepackage[T1]{fontenc}
\usepackage[utf8]{inputenc}
\usepackage{ae,aecompl}
\usepackage{dutchcal}
\usepackage{graphicx}	
\usepackage{amsmath}	
\usepackage{amssymb}	
\usepackage{lscape}
\usepackage{caption}

\DeclareMathOperator*{\argmin}{arg\,min}

\usepackage{etoolbox}
\makeatletter
\patchcmd\@combinedblfloats{\box\@outputbox}{\unvbox\@outputbox}{}{%
    \errmessage{\noexpand\@combinedblfloats could not be patched}%
}%
\makeatother

\title[Deep Learning for Star Cluster Classification]{Deep Transfer Learning for Star Cluster Classification: I. Application to the PHANGS-HST Survey}

\author[Wei Wei et al.]
{Wei Wei,$^{1,2}$\thanks{Contact e-mail: \href{mailto:weiw2@illinois.edu}{weiw2@illinois.edu}}
E. A. Huerta,$^{1,3}$
Bradley C. Whitmore,$^{4}$
Janice C. Lee,$^{5}$
\newauthor
Stephen Hannon,$^{6}$
Rupali~Chandar,$^{7}$
Daniel~A.~Dale,$^{8}$
Kirsten~L.~Larson,$^{5}$
\newauthor
David~A.~Thilker,$^{9}$
Leonardo~Ubeda,$^{4}$
M\'ed\'eric~Boquien,$^{10}$
\newauthor
M\'{e}lanie Chevance,$^{11}$
J.~M.~Diederik Kruijssen,$^{11}$
Andreas Schruba$^{12}$,
\newauthor
Guillermo Blanc$^{13,14,15}$,
Enrico Congiu$^{16,13}$
\\
$^{1}$NCSA, University of Illinois at Urbana-Champaign, Urbana, Illinois 61801, USA\\
$^{2}$Department of Physics, University of Illinois at Urbana-Champaign, Urbana, Illinois 61801, USA\\
$^{3}$Department of Astronomy, University of Illinois at Urbana-Champaign, Urbana, Illinois 61801, USA\\
$^{4}$Space Telescope Science Institute, 3700 San Martin Drive, Baltimore, MD, USA\\
$^{5}$Caltech/IPAC, California Institute of Technology, Pasadena, CA, USA\\
$^{6}$Department of Physics and Astronomy, University of California, Riverside, CA, USA\\
$^{7}$Department of Physics and Astronomy, University of Toledo, Toledo, OH USA\\
$^{8}$Department of Physics and Astronomy, University of Wyoming, Laramie, WY, USA\\
$^{9}$Department of Physics and Astronomy, The Johns Hopkins University, Baltimore, MD, USA\\
$^{10}$Unidad de Astronom\'ia, Universidad de Antofagasta, Antofagasta, Chile\\
$^{11}$Astronomisches Rechen-Institut, Zentrum f\"ur Astronomie der Universit\"at Heidelberg, Heidelberg, Germany\\
$^{12}$Max-Planck-Institut f\"ur extraterrestrische Physik, Garching, Germany\\
$^{13}$Observatories of the Carnegie Institution for Science, Pasadena, CA, USA\\
$^{14}$Departamento de Astronomía, Universidad de Chile, Las Condes, Santiago, Chile\\
$^{15}$Centro de Astrofísica y Tecnologías Afines (CATA), Las Condes, Santiago, Chile\\
$^{16}$Las Campanas Observatory, La Serena, Chile\\
}

\date{Accepted XXX. Received YYY; in original form ZZZ}

\pubyear{2019}

\begin{document}
\label{firstpage}
\pagerange{\pageref{firstpage}--\pageref{lastpage}}
\maketitle

\begin{abstract}
 We present the results of a proof-of-concept experiment which demonstrates that deep learning can successfully be used for production-scale classification of compact star clusters detected in HST UV-optical imaging of nearby spiral galaxies (\(D\lesssim20\,\textrm{Mpc}\)) in the PHANGS-HST survey. 
Given the relatively small nature of existing, human-labelled star cluster samples, we transfer the knowledge of state-of-the-art neural network models for real-object recognition to classify star clusters candidates into four morphological classes. 
 We perform a series of experiments to determine the dependence of classification performance on: neural network architecture (ResNet18 and VGG19-BN); training data sets curated by either a single expert or three astronomers; and the size of the images used for training. We find that the overall classification accuracies are not significantly affected by these choices. The networks are used to classify star cluster candidates in the PHANGS-HST galaxy NGC 1559, which was not included in the training samples.  The resulting prediction accuracies are 70\%, 40\%, 40-50\%, 50-70\% for class  1,  2,  3  star  clusters, and class 4 non-clusters respectively. This performance is competitive  with consistency achieved in previously published human and automated quantitative classification of star  cluster candidate  samples (70-80\%, 40-50\%, 40-50\%, and 60-70\%). The methods introduced herein lay the foundations to automate classification for star clusters at scale, and exhibit the need to prepare a standardized dataset of human-labelled star cluster classifications, agreed upon by a full  range  of  experts  in  the  field, to further improve the performance of the networks introduced in this study.
\end{abstract}



\section{Introduction} 
\label{sec:intro}

Human visual classification of electromagnetic signals from astronomical sources is a core task in observational research with a long established history~\citep{cannon1912, cannon1918, E_H_1926_ApJ,1936_book_E_H,d_Vau_1963ApJS}. It has been an essential means by which progress has been made in understanding the formation and evolution of structures from stars to galaxies.  However, in the modern era of ``Big Data" in Astronomy, with unprecedented growth in electromagnetic survey area, field of view, sensitivity, resolution, wavelength coverage, cadence, and transient alert production, it has become apparent that human classification is no longer scalable~\citep{DES:2016MNRAS,lsstbook}. This realization has motivated the use of machine learning techniques to automate image classification~\citep{Ball_2008_ApJ,Banerji_2010_MNRAS,Carrasco_2013_MNRAS,ml_book_2017,Kamdar_2016,Kim_Bru_MNRAS_2017}. Some of these machine learning algorithms have been integrated into widely-used methods for image processing, such as the neural networks trained for star/galaxy separation in the automated source detection and photometry software \texttt{\textbf{SE}XTRACTOR}~\citep{Bertin_1996AAS}. Other applications of machine learning for image classification include the use of so-called decision trees~\citep{weir_1995AJ,Suchkov_2005AJ,Ball_2006ApJB,Vasconcellos_2011AJ,Sevilla_2015ACS} and support vector machines~\citep{Fadely_2012ApJF,Solarz_2017AAS,Male_2013AAM}.

Visual object recognition has also been a core research activity in the computer science community. For instance, the PASCAL VOC challenge was initiated to develop software to accurately classify about 20,000 images divided into twenty object classes~\citep{pascalvoc}. Over the last decade deep learning algorithms have rapidly evolved to become the state-of-the-art signal-processing tools for computer vision, to the point of surpassing human performance. The success of deep learning algorithms for image classification can be broadly attributed to the combination of increasing processing speed and the availability of very large datasets for training; i.e., Graphics Processing Units (GPUs) to train, validate and test neural network models; and curation of high-quality, human-labeled datasets, such as the \texttt{ImageNet} dataset~\citep{imagenet_cvpr09}, which has over 14 million images divided into more than 1000 object categories. 

The \texttt{ImageNet} Large Scale Visual Recognition Challenge \citep{ILSVRC15} has driven the development of deep learning models that have achieved breakthroughs for image classification. In 2012, the network architecture \texttt{AlexNet}~\citep{krizhevsky2012imagenet} achieved a \(\sim50\%\) reduction in error rate in the  \texttt{ImageNet} challenge---a remarkable feat at that time that relied on the use of GPUs for the training of the model, data augmentation (image translations, horizontal reflections and mean subtraction), as well as other novel algorithm improvements that are at the core of state-of-the-art neural network models today, e.g., using successive convolution and pooling layers followed by fully-connected layers at the end of the neural network architecture. 

Within the next two years, the architectures \texttt{VGGNet}~\citep{vggnet} and \texttt{GoogLeNet}~\citep{googlenet} continued to improve the discriminative power of deep learning algorithms for image classification using deeper and wider neural network models, and innovating data augmentation techniques such as scale jittering. Furthermore, \texttt{GoogLeNet} provided the means to  
further improve image classification analysis by introducing multi-scale processing, i.e., allowing the neural network model to recover local features through smaller convolutions, and abstract features with larger convolutions. In 2015, the \texttt{ResNet}~\citep{resnet} model was the first architecture to surpass human performance on the \texttt{ImageNet} challenge. In addition to this milestone in computer vision, \texttt{ResNet} was also used to demonstrate that a naive stacking of layers does not guarantee enhanced performance in ultra deep neural network models, and may actually lead to sub-optimal performance for image classification.

In view of the aforementioned accomplishments, research in deep learning for image classification has become a booming enterprise in science and technology. This vigorous program has led to innovative ways to leverage state-of-the-art neural network models to classify disparate datasets. This approach is required because most applications of deep learning for image classification rely on supervised learning.  That is, neural network models are trained using large datasets of labelled data, such as the \texttt{ImageNet} dataset.  In astronomical research, to enable the morphological classification of galaxies, the deep neural network model developed by \citep{Dieleman15} was trained on $\sim$55,000 galaxy images, each with 40-50 human classifications from the Galaxy Zoo 2 \citep{willett13} online crowdsourcing project.  This model was developed for the Galaxy Challenge competition in 2013-14 on the Kaggle platform, and took first place out of 326 entries.  \color{black} Given that datasets of that nature are challenging to obtain, deep ``transfer'' learning has provided the means to classify entirely new datasets {\it by fine-tuning a pre-trained neural network model} with the \texttt{ImageNet} dataset.\footnote{A brief overview of transfer learning is presented in Appendix~\ref{deeptransfer}.} 

While deep transfer learning was initially explored to classify datasets that were of similar nature to those used to train state-of-the-art neural network models, the first application of deep transfer learning of a pre-trained \texttt{ImageNet} neural network model to classify small datasets of entirely different nature was presented in~\cite{dglitcha:2017,dgNIPS}, where a variety of neural network models were used to report state-of-the-art image classification accuracy of noise anomalies in gravitational wave data. That study triggered a variety of applications of pre-trained \texttt{ImageNet} deep learning algorithms to classify images of galactic mergers~\citep{dtl_galaxy_merger}, and  galaxies~\citep{Asad:2018,barchi_2019,Dom:2018D}, to mention a few examples. 

Building upon these recent successful applications of deep transfer learning for image classification in physics and astronomy, in this paper we demonstrate that deep transfer learning provides the means to classify images of compact star clusters in nearby galaxies obtained with the Hubble Space Telescope (HST).  We show that this approach yields classification accuracies on par with work performed by humans, and has the potential to \color{black} outperform humans and traditional machine learning.   A major motivation of this work is to determine whether these deep transfer learning techniques can be used to automate production-scale classification of candidate star clusters in data from the Cycle 26 \texttt{HST-PHANGS} (Physics at High Angular Resolution in Nearby GalaxieS\footnote{\url{www.phangs.org}}) Survey (PI: J.C. Lee, GO-15654) for which observations commenced in April 2019.  \texttt{HST-PHANGS} is anticipated to yield several tens of thousands of star cluster candidates for classification, only about a half of which will be true clusters.  Encoding classification systems in neural networks will also improve the consistency of the classifications, and reduce the implicit impacts of subjectivity and subtle differences in classification systems adopted by different individuals (i.e., it can reduce both random and systematic errors in the classifcations).\color{black}

This paper is organized as follows.  In Section~\ref{sec:class}, we summarize the objectives of star cluster classification, and describe the current classification system, which we employ in this paper.  A review of the consistency between human classifications across prior studies is provided to establish the accuracy level to be achieved or surpassed by deep learning in this initial proof-of-concept experiment.  In Section~\ref{sec:met}, we describe the imaging data and classifications used to train our neural network (NN) models, and then provide an overview of the NN models employed in this work.  We report our results in Section~\ref{sec:res}.  We conclude in Section~\ref{sec:end} with a summary of the results and next steps for future work. 

\section{Classification of Compact Star Clusters in Nearby Galaxies}
\label{sec:class}
The objects of interest in this study are compact star clusters and stellar associations in galaxies at distances between 4 Mpc to 20 Mpc.  The physical sizes of compact clusters are characterized by effective radii between 0.5pc to about 10pc~\citep{pz10,Ryon_2017}.  \cite{ryon14} report that the distribution of effective radii of young ($\lesssim$10 Myr), massive compact star clusters peaks between 2-3 pc based on HST LEGUS observations of NGC1313 (D$\sim$4 Mpc) and NGC628 (D$\sim$10 Mpc).   Hence, only with the resolution of HST\footnote{The WFC3/UVIS point source function FWHM is 0\farcs067 at 5000\AA.} 
can such objects be distinguished from individual stars and separated from other star clusters in galaxies beyond the Local Group. \footnote{We note that for a high signal-to-noise cluster it is possible to measure the broadening  of the image (and hence the size of the source) to a fraction of the FWHM of the PSF of a star. The FWHM of a star using WFC3 is about 1.8 pix (1.3 pc at D=4 Mpc, and 6.4 pc at 20 Mpc). A significant amount of testing has been done on ACS and WFC3 images using software like ISHAPE (Larson1999), and much published work (including Chandar et al. 2017, Ryon et al. 2017) has confirmed that this broadening can be measured down to about 0.2 pixels, corresponding to size limits of $\sim$0.3 pc, $\sim$0.6 pc at distances of 5 Mpc, 10 Mpc. Extending to 15 and 20 Mpc, the upper end of distance range covered by the PHANGS survey, the cluster size limits are 0.8 and 1.1 pc. Per the ISHAPE manual, at 5 Mpc, this is calculated as: 0.2 pix * 0.04 (arcsec/pix)* 24 pc/arcsec * 1.48 = 0.28 pc (where 1.48 is a conversion factor given in the ISHAPE manual when assuming a King profile specifically). Hence, if the peak sizes for clusters are in the 2-3 pc range, the vast majority of cluster will be resolved for most of the galaxies in PHANGS-HST.} 
The sizes of stellar associations, which dominate the young stellar population, span a wider range with sizes from a few pc to $\sim$100 pc ~\citep{pz10,gouliermis18}.\color{black}

Early attempts at classifying clusters in external galaxies with HST imaging focused mainly on old globular clusters, for example, the swarm of thousands of globular clusters around the central elliptical galaxy in
the Virgo Cluster, M87 ~\citep{whitmore95}. This was a fairly straightforward process since the background was smooth and the clusters were well separated.  With the discovery of super star clusters in merging galaxies \citep[e.g,][]{holtzman92}, the enterprise of the identification and study of clusters in star-forming galaxies using HST began, despite the fact that crowding and variable backgrounds in such galaxies make the process far more challenging.  Studies of normal spiral galaxies pushed the limits to fainter and more common clusters \citep[e.g,][]{larsen02, chandar10b}.  In all these early studies, the primary objective was to distinguish true clusters from individual stars and image artifacts, and there were essentially no attempts to further segregate the clusters into different classes.

An exception, and one of the first attempts at a more detailed classification, was performed by \cite{schweizer96}, who defined 9 object types and then grouped them into two classes: candidate globular clusters and extended stellar associations. More recently, \cite{bastian12a}, who studied clusters using HST imaging of the M83 galaxy, classified star clusters as either symmetric or asymmetric. Their analysis retained only symmetric clusters, which they posited were more likely to be gravitationally bound. Following this work, many studies in the field, most notably the Legacy ExtraGalactic UV Survey (\texttt{LEGUS})~\citep{legus} began differentiating clusters into two or three different categories, so that they could be studied separately or together depending on the goals of the project \citep[see also the review by][and their discussion of ``exclusive" versus ``inclusive" cluster catalogs]{krumholz18}.  

The \texttt{LEGUS} project also employed machine learning techniques for some of their cluster classification work \cite{messa18, grasha19}. This pioneering work will be discussed in Section 5.

In \texttt{LEGUS}, cluster candidates are sorted into four classes as follows~\citep{adamo17, cook19}:

\begin{itemize}
\item Class 1: compact, symmetric, single central peak, radial profile more extended relative to point source
\item Class 2: compact, asymmetric or non-circular (e.g., elongated), single central peak

\item Class 3: asymmetric, multiple peaks, sometimes superimposed on diffuse extended source
\item Class 4: not a star cluster (image artifacts, background galaxies, pairs and multiple stars in crowded regions, stars)
\end{itemize}

\begin{figure*}
    \centering
    \includegraphics[width=0.85\textwidth]{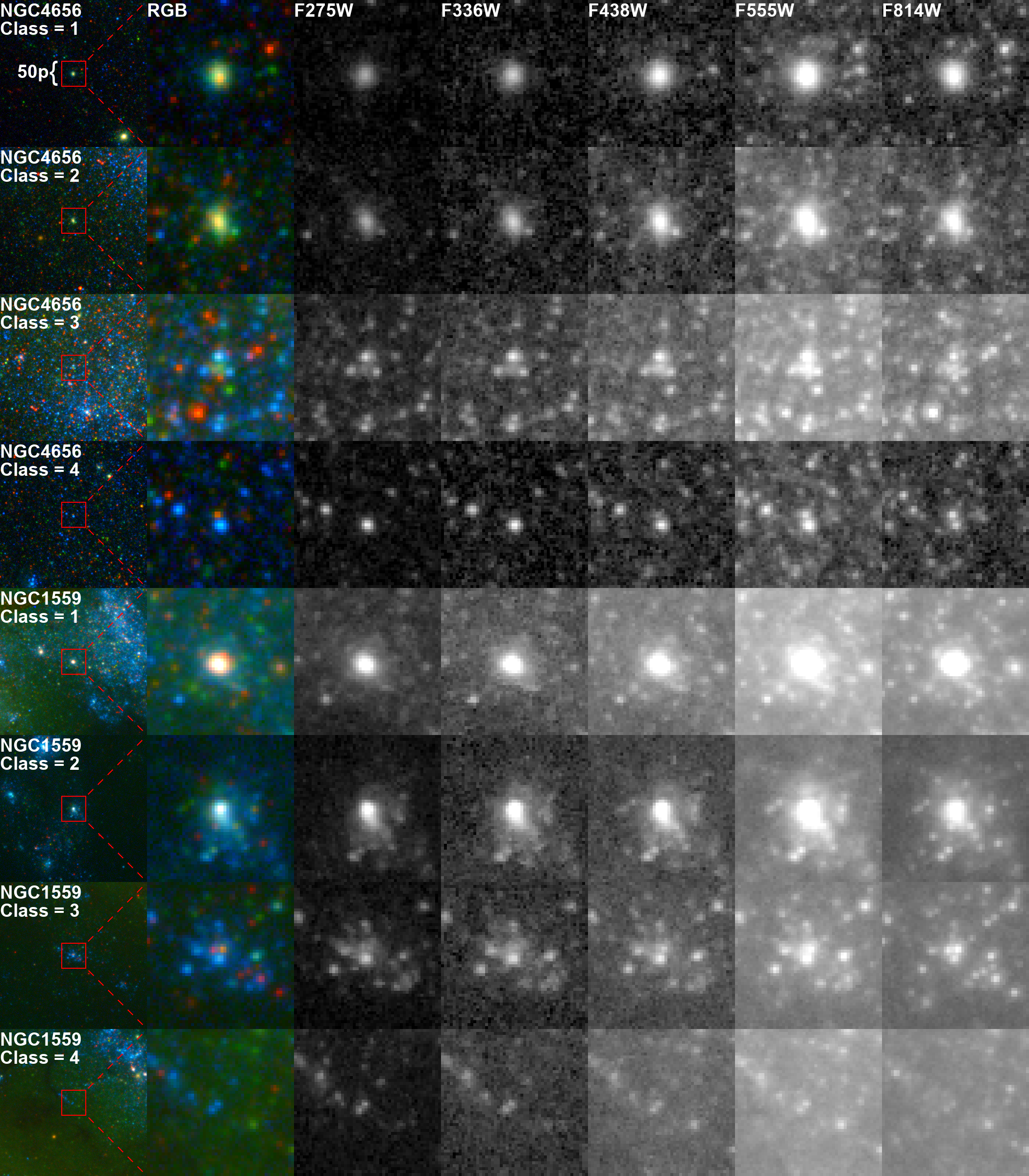}
    \caption{Examples of each of the four cluster classifications illustrated with HST/WFC3 imaging. The top four rows show star clusters from NGC 4656, which are part of the training set, while the bottom four rows show clusters from recent PHANGS-HST observations of the spiral galaxy NGC 1559, which form our proof-of-concept test sample, and are not used for training. The first two columns show false-color RGB images for context: the first column displays a 299p x 299p RGB image (R = F814W, G = F438W + F555W, B = F275W + F336W) and the second column shows only the center 50p x 50p of the RGB image (184pc x 184pc for NGC1559, for example).  The center 50p x 50p of individual NUV-U-B-V-I HST images, which are used as input to the pre-trained neural network models for further training (tuning) and evaluation, are shown in grayscale in the last 5 columns (from left to right, 50p x 50p images taken with filters F275W, F336W, F438W, F555W, and F814W).  We also experiment with 25p x 25p and 100p x 100p images, as discussed in Sections 3 and 4.}
    \label{fig:clusterclass_illustration}

\end{figure*}

We adopt the same classification system for this paper.  In general, we refer to class 1, 2, and 3 as ``compact symmetric cluster," ``compact asymmetric cluster," and ``compact association" respectively.  Examples of objects in each of these classes are shown in Figure~\ref{fig:clusterclass_illustration}.

\subsection{Consistency among  Classifications}
\color{black}
\label{sec:c_human}
The stated goal of the current work is to provide cluster classifications via deep transfer learning models that achieve accuracy levels at least 
 
 as good as other star cluster classifications in the literature, both by human visual inspection and by application of quantitative selection criteria. \color{black}
In this section we establish this ``accuracy" level, which we define as the consistency between different classifications for the same cluster populations as reported in the literature, as well as relative to classifications homogeneously performed by one of us (Bradley C.~Whitmore, hereafter BCW.). 

A first look at the overall consistency between the clusters cataloged by different studies, but based on the same data and same limiting magnitude, is provided by the work on M83 by \cite{bastian12a, whitmore14, chandar14}.  Comparisons reported in those papers show that about $\sim$70\% of the clusters are in common between the studies. Later, \cite{adamo17} performed a similar comparison for the spiral galaxy NGC 628 for the catalogs from \texttt{LEGUS} and \cite{whitmore14}, and finds an overlap

of $\sim$75\%.  Finally, the LEGUS study of M51 by \cite{messa18} find 

an overlap
of 73\% in common with a study by \cite{chandar16}.

These results are not based only upon detailed analysis of human-vs-human cluster classifications for individual objects; they are statistical measures of overlap between samples where a mix of human classification/identification, and automated star/cluster separation based on the concentration index (i.e., the difference in magnitude in a 1 pixel vs. 3 pixel radius) were used across the studies.

\begin{figure}
    \centering
    \includegraphics[width=\linewidth]{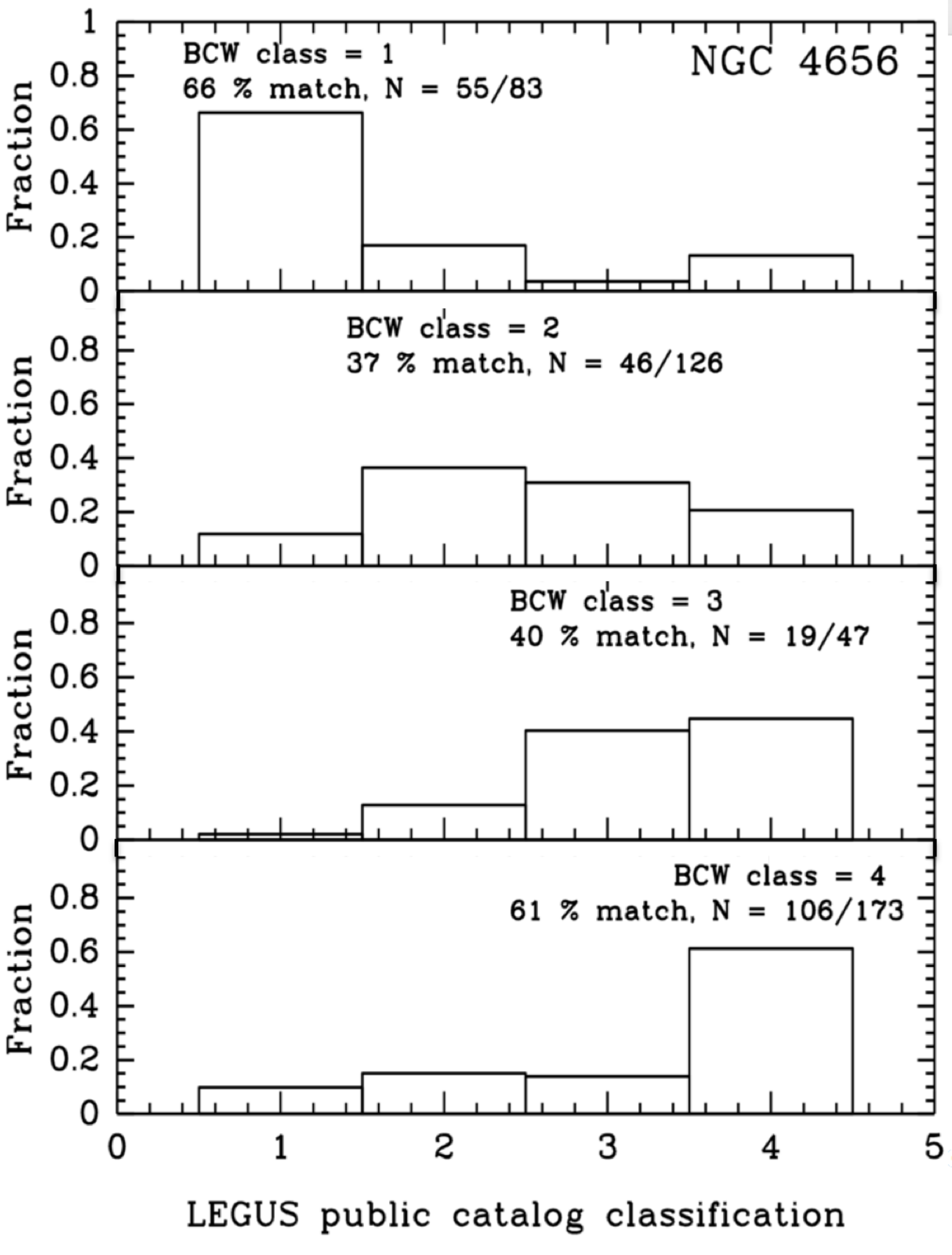}
    \caption{Comparisons between star cluster candidate classifications made by BCW and the mode of classifications made by three other \texttt{LEGUS} team members (trained by BCW, A. Adamo, and H. Kim) provided in the LEGUS public star cluster catalog for NGC 4656.  Each panel shows the distribution of classifications given in the \texttt{LEGUS} catalog for BCW labelled class 1 (top, symmetric compact clusters), class 2 (upper middle, asymmetric compact clusters), class 3 (lower middle, compact associations) and class 4 (bottom, non-clusters) objects. }
    \label{fig:ngc4656_classcompare}
\end{figure}

To more directly evaluate human-vs-human cluster classifications alone we 
 start with a comparison of the NGC 3351 cluster catalog from the LEGUS sample (performed by BCW and team member Sean Linden, who was trained by BCW) with a new version of the NGC3351 cluster catalog independently constructed by PHANGS-HST\footnote{PHANGS-HST has expanded imaging coverage of NGC3351 to produce greater overlap with PHANGS-ALMA CO observations of the galaxy, and is developing new star cluster catalogs for the fields. See Section~\ref{sec:catalog_construction} for an overview of the catalog construction.}  (performed by BCW alone). This might be viewed as a test of the consistency that might be expected if the same (or very similar) classifiers return to the same data set after a passage of several years.
We find a 80 \% agreement between category 1 objects, 
53 \% for category 2, 56 \% for category 3. If we combine category 1 and 2 objects (which is what many authors do for their analysis), the agreement is 88 \%.

We next compare classifications assigned by BCW for NGC 4656 to those provided in the \texttt{LEGUS} public cluster catalog, which provides the mode of classifications made by three other \texttt{LEGUS} team members (trained by BCW, A. Adamo, and H. Kim).  Results are shown in Figure~\ref{fig:ngc4656_classcompare}.

If we
combine only the class $1 + 2$ clusters (to exclude compact associations which has a higher rate of confusion with class 4 non-clusters), the total match fraction is 67\%.
For the individual classes, the consistency of the assignments vary from 66\%, 37\%, 40\%, 61\% for class 1, 2, 3, and 4, respectively. 
 Hence, the  agreement for the BCW classifications versus the mode of classifications from three LEGUS team members for NGC 4656 are slightly lower than the comparisons between BCW and BCW (and Linden) for NGC 3351.
Other galaxies where a similar comparison has been made between the BCW classifications and LEGUS 3-person (``consensus'') classifications (i.e., NGC 4242, NGC 4395N, and M51) result in similar numbers.

In summary, comparing between a wide range of different cluster classification methods, but for the same data sets, we find typical agreements in the range 40 \% (e.g., when comparing class 2 or class 3 objects alone)  to 90 \% (e.g. when combining class 1 + 2  for repeat classifications of 
 
cluster catalogs by the same, or very similar, classifiers).  For the individual classes, the ``accuracy" levels that we adopt to be achieved or surpassed for our deep learning studies proof-of-concept demonstration are 70-80\%, 40-50\%, 40-50\%, and 60-70\% for class 1, 2, 3, 4 objects respectively.

\color{black}

\begin{table*}
\begin{tabular}{llllll}
\hline
\hline
Field            & D (Mpc) & Class 1 & Class 2 & Class 3 & Class 4 \\
\hline
NGC3351$^1$      & 10.0     & 118     & 80      & 95      & 325     \\
NGC3627          & 10.1     & 403     & 175     & 164     & 837     \\
NGC4242$^1$      & 5.8      & 117     & 60      & 14      & 42      \\
NGC4395N$^2$     & 4.3      & 8       & 19      & 21      & 20      \\
NGC4449          & 4.31     & 120     & 261     & 213     & 0       \\
NGC45$^1$        & 6.61     & 45      & 52      & 20      & 43      \\
NGC4656$^2$      & 5.5      & 83      & 125     & 47      & 173     \\
NGC5457C         & 6.7      & 287     & 108     & 81      & 436     \\
NGC5474$^1$      & 6.8      & 48      & 95      & 34      & 144     \\
NGC6744N         & 7.1      & 164     & 143     & 58      & 210     \\
\hline
\textbf{Total}           &          & 1393    & 1118    & 747     & 2230    \\
\textbf{N $\geq$ 4} &          & 1271    & 1013    & 738     & 2125
\\
\hline
\hline
\end{tabular}
\caption{Number of sources in each of the ten HST LEGUS fields which have been primarily classified by BCW. The number in each of morphological classes described in Section~\ref{sec:class} is given.  The total number of clusters with detection in at least four filters (a requirement for inclusion in the training and testing) are given in the last row of the table.  80\% of the latter (randomly selected) are used for training, and the remaining 20\% are reserved for validation testing.  Distances compiled by \citep{legus} are listed. 
\newline$^1$ Classification primarily determined by BCW are available in the public release of the LEGUS cluster catalogs. 
\newline$^2$ Independent classifications determined by BCW for fields for which LEGUS consensus classifications are available through the LEGUS public archive (Table~\ref{tab:leguspubliccatalogs}).}
\label{tab:sample}
\end{table*}
\begin{table*}
\begin{tabular}{llllll}
\hline
\hline
Field    & D (Mpc)        & Class 1 & Class 2 & Class 3 & Class 4 \\
\hline
NGC1313E & 4.39           & 42      & 95      & 122     & 386     \\
NGC1313W & 4.39           & 85      & 191     & 210     & 373     \\
NGC1433  & 8.3            & 51      & 61      & 56      & 138     \\
NGC1566  & 18.0           & 258     & 214     & 261     & 328     \\
NGC1705  & 5.1            & 16      & 13      & 13      & 54      \\
NGC3344  & 7.0            & 119     & 118     & 159     & 161     \\
NGC3738  & 4.9            & 49      & 93      & 86      & 214     \\
NGC4656  & 5.5            & 93      & 91      & 78      & 169     \\
M51      & 7.66           & 363     & 502     & 365     & 1261    \\
NGC5253  & 3.15           & 20      & 37      & 23      & 154     \\
NGC628C  & 9.9            & 334     & 357     & 326     & 542     \\
NGC628E  & 9.9            & 92      & 80      & 87      & 122     \\
NGC6503  & 5.27           & 71      & 96      & 131     & 172     \\
NGC7793E & 3.44           & 32      & 76      & 83      & 62      \\
NGC7793W & 3.44           & 51      & 84      & 86      & 78      \\
IC4247   & 5.1            & 1       & 4       & 3       & 37      \\
IC559    & 5.3            & 9       & 12      & 4       & 18      \\
NGC4395N & 4.3            & 8       & 12      & 19      & 19      \\
NGC4395S & 4.3            & 31      & 64      & 42      & 31      \\
NGC5238  & 4.51           & 4       & 4       & 1       & 9       \\
NGC5477  & 6.4            & 5       & 9       & 9       & 49      \\
UGC1249  & 6.9            & 13      & 35      & 40      & 133     \\
UGC4305  & 3.05           & 16      & 29      & 40      & 147     \\
UGC4459  & 3.66           & 2       & 5       & 3       & 20      \\
UGC5139  & 3.98           & 2       & 7       & 7       & 23      \\
UGC685   & 4.83           & 7       & 4       & 3       & 6       \\
UGC695   & 10.9           & 4       & 7       & 6       & 94      \\
UGC7408  & 6.7            & 19      & 16      & 11      & 32      \\
UGCA281  & 5.9            & 2       & 9       & 4       & 34      \\
\hline
\textbf{Total}   &       & 1799    & 2325    & 2278    & 4866    \\
\textbf{N $\geq$ 4}       &       & 1795    & 2315    & 2265    & 4841    \\
\hline
\hline
\end{tabular}
\caption{Same as Table~\ref{tab:sample}, but for the 29 HST LEGUS fields which have been classified by three people, and have star cluster catalogs available through the LEGUS public archive.  The number in each of the morphological classes, as determined by the mode of these three people's classifications, is given.}
\label{tab:leguspubliccatalogs}
\end{table*}
\begin{table*}
\begin{tabular}{llllll}
\hline
\hline
Field   & D (Mpc) & Class 1 & Class 2 & Class 3 & Class 4 \\
\hline
NGC1559 & 19.0     & 302     & 252     & 162     & 710 
\\
\hline
\hline
\end{tabular}
\caption{Number of sources in the PHANGS-HST observation of NGC 1559 which have been classified by BCW. This cluster sample is used to test the neural networks trained as described in Section~\ref{sec:res_1} as a proof-of-concept demonstration for production scale classification of PHANGS-HST compact clusters and associations.}
\label{tab:ncg1559}
\end{table*}

\section{Data and Methods} 
\label{sec:met}

In this section we describe the data sets used to train, validate and test our deep learning algorithms, and give an overview of the neural network models used. We approach this initial work as a proof of concept demonstration, with the intention of performing further optimization and more detailed tests in future work.
 
 \subsection{Star Cluster Catalogs}
 \label{sec:catalog_construction}
 
A key point is that the training and testing of the neural networks presented here are based on a pre-selected sample of cluster candidates where a large fraction of unresolved (point) sources have been first discarded.  In past work, such candidate samples have served as the starting point for visual classification by humans to remove remaining interlopers, and to characterize the morphologies of verified clusters as described above.  The construction and selection methodolgy for cluster candidate samples used here follow most of the procedures adopted for the LEGUS project \citep{calzetti15} as described in~\cite{adamo17}.  

To briefly review, the procedure includes detection using the SExtractor program~\citep{sex96} on a white light image; filtering out most stars by requiring the concentration index\footnote{(CI = difference in magnitude between an aperture with 1 or 3 pixels)} to be greater than a value determined based on training set of isolated point sources and clusters for each galaxy; requiring detections with photometric errors less than 0.3 mag in at least 4 filters; and selecting objects brighter than -6 mag in F555W (total Vega magnitude).  Again, this results in a cluster candidate list which is then examined visually to remove artifacts (e.g., close pairs of stars, saturated stars and diffraction spikes, background galaxies, etc.). The primary tool used for the visual classification is the IMEXAMINE task in IRAF. 
See Figure 3 in~\cite{adamo17} for a graphic description of the use of IMEXAMINE and the classification into four categories.

For most of the LEGUS star cluster catalogs which have been publicly released through the MAST archive, \footnote{\url{https://archive.stsci.edu/prepds/legus/dataproducts-public.html}} classifications are performed by three different team members and the mode is recorded as the final consensus value (i.e., the 29 fields in Table~\ref{tab:leguspubliccatalogs}).  The LEGUS classifiers, were trained by BCW, A. Adamo, and H. Kim.  For an additional 8 fields, classifications were performed primarily by a single team member, i.e., coauthor BCW.\footnote{S. Linden who was trained by BCW, assisted in classifications for sources in NGC 3351, NGC 3627, and NGC 5457)}  As of July 2019, classifications for 4 of the 8 HST fields primarily inspected by BCW  
are available from the \texttt{LEGUS} public archive (Table~\ref{tab:sample}).  BCW also independently classified two fields with LEGUS consensus classifications to enable consistency checks (e.g., Figure~\ref{fig:ngc4656_classcompare}),
bringing the total to 10 galaxies in the sample with BCW classifications.

The construction of a preliminary cluster catalog for the first galaxy observed in the PHANGS-HST program NGC 1559 generally follow the methods used for LEGUS.  The primary differences are that a F555W image was used instead of a white light image (which is more prone to small differences in alignment of different filters and the presence of very close pairs of stars with different colors),

and a false-color image from the Hubble Legacy Archive \citep{hla} was simultaneously examined to help classify the clusters.  A magnitude limit of -7.5 in the V band was used for NGC 1559, reflecting its larger distance (19 Mpc: A. Reiss, private communication) relative to the average distance of the LEGUS galaxies.  A detailed presentation of the PHANGS-HST star cluster and association candidate selection methods will be provided in the PHANGS-HST survey paper (Lee et al. 2020) and catalog papers (e.g., Whitmore et al. 2020, Thilker et al. 2020, Larson et al. 2020).

\color{black}
\subsection{Image data \& curation}

As input for the neural network training, we use postage stamps extracted from HST imaging taken in five broadband filters. 
Sample postage stamps are presented in the last five columns of Figure~\ref{fig:clusterclass_illustration}.  

LEGUS obtained HST observations with WFC3 in 2013-2014 (GO-13364; PI Calzetti), and combined those data with ACS data taken in previous cycles by other programs to provide \texttt{NUV-U-B-V-I} coverage for a sample of 50 galaxies with 63 fields.

PHANGS-HST (GO-15654; PI Lee) began observations on April 6, 2019 and is also obtaining observations with similar exposure times in the NUV-U-B-V-I filters.  The first galaxy to be observed is NGC 1559.

Bearing in mind that the neural network models used in this study (i.e., \texttt{VGG19-BN} and \texttt{ResNet18}; see next section) were pre-trained with the \texttt{ImageNet} dataset, in which images are resized to \(299\times 299\times3\), we follow best coding practices of neural network training, and curate our datasets so that star cluster images have size  \(299\times 299\) pixels.

However, given that star clusters subtend only about several to a dozen HST WFC3 pixels, we focus the training on a small area (see Figure~\ref{fig:clusterclass_illustration}).  

We first extract regions of 50 x 50 HST/WFC3 pixels centered on the star cluster candidate, which are then resized to fit in an 299 x 299 pixel area for the training.  
With WFC3's pixel size of 0.04 arcseconds, each region corresponds to a physical width between $\sim$40-100pc for our sample of galaxies. To test whether the size of the cropped HST image influences the accuracy, we also extract regions which are half and twice as large as 50 HST/WFC3 pixels across.

Procedurally, from the HST mosaics, a .fits image ``postage stamp" centered on each target cluster is cropped from each of the NUV-U-B-V-I bands. 

The five resultant stamps for each cluster candidate are then stored in individual header data units (HDUs) within a single MEF file. We note that if there was no observation of the cluster in one of the filters, all pixel values for that particular filter's postage stamp were set to zero. If there was no observation in more than one filter, the cluster was removed from our sample, consistent with the candidate selection criteria.

\subsection{Neural network models}
\label{drl}
The available star cluster data sets are small compared to the datasets used to successfully train state-of-the-art neural network models for image classification.  Thus, we use two neural network models, \texttt{VGG19}~\citep{simonyan2014very} with batch normalization (\texttt{VGG19-BN}) and \texttt{ResNet18}~\citep{he2016deep}, pre-trained with the \texttt{ImageNet} dataset (see Section~\ref{sec:intro}), and then use deep transfer learning\footnote{A brief overview of transfer learning is presented in Appendix~\ref{deeptransfer}.} to leverage the knowledge of these models to classify real-object images to our task at hand, namely, the morphological classification of star clusters.

Regarding batch normalization for VGG19: the weights of each layer in a neural network model change throughout the training phase, which implies that the activations of each layer will also change. Given that the activations of any given layer are the inputs to the subsequent layer, this means that the input distribution changes at every step. This is far from ideal because it forces each intermediate layer to continuously adapt to changing inputs. Batch normalization is used to ameliorate this problem by normalizing the activations of each layer. In practice this is accomplished by adding two trainable parameters to each layer, so the normalized output is multiplied by a standard deviation parameter, and then shifted by a mean parameter. With this approach only two parameters are changed for each activation, as opposed to losing the stability of the network by changing all the weights. It is expected that through this method each layer will learn on a more stable distribution of inputs, which may accelerate the training stage.

Both neural network architectures, \texttt{VGG19-BN} and \texttt{ResNet18} have 3 input channels. However, since the star cluster candidates have images taken in 5 broadband filters,

we concatenate two copies of the same neural network architecture. The merged neural networks have 6 input channels in total, so we set the input to the last channel to be constant zeros. 
We also apply one more matrix multiplication and an element-wise softmax function (see Appendix~\ref{sec:stats})~\citep{Goodfellow-et-al-2016} to make sure that for each 

candidate cluster the output is a vector of size 4, representing the probability distribution over the 4 classes under consideration. We choose this particular combination given its simplicity and its expected performance for image classification.

We use the pre-trained weights, except those for the last layers, of \texttt{VGG19-BN} and \texttt{ResNet18} provided by \texttt{PyTorch}~\citep{paszke2017automatic} as the initial values for the weights in our models. The weights for the last layers in \texttt{VGG19-BN} and \texttt{ResNet18} and the last fully connected layers are randomly initialized. We use cross-entropy as the loss function\footnote{A loss function is used to evaluate and diagnose model optimization during training. The penalty for errors in the cross-entropy loss function is logarithmic, i.e., large errors are more strongly penalized.} and \texttt{Adam}~\citep{kingma2014adam} for optimization. The learning rate is set to $10^{-4}$. The batch size for \texttt{ResNet18} is 32, and for \texttt{VGG19-BN} is 16. 

Batch size and batch normalization refer to two distinct concepts. One epoch corresponds to all the training examples being passed both forward and backward through the neural network only once, while the batch size is the number of training examples in one forward/backward pass. For instance, we may divide a training data set of 100 images into 4 batches, so that the batch size is 25 sample images, and 4 iterations will complete one epoch. On the other hand, batch normalization is a technique used to improve the stability of the learning algorithms. The details are described in  Appendix \ref{batchnorm}.

Finally, following deep learning best practices, we quantify the variance in classification performance of our models by training them ten times independently and then presenting the mean accuracies and the corresponding standard deviations. We also compute the Shannon entropy ~\cite{shannon1948mathematical} of the output distribution over the four star cluster classes to quantify the uncertainty in each individual neural network model's prediction. 

\subsection{Training Experiments}

 We perform a series of experiments to test how the accuracy of the neural network model for predicting the morphological classification of candidate star clusters depends on the following characteristics of the training sample:
\begin{enumerate}
	\item origin of classifications: primarily classified by BCW (Table~\ref{tab:sample}) or the mode of 3 LEGUS classifiers (Table~\ref{tab:leguspubliccatalogs})
	\item size of images used for training: 25p x 25p, 50p x 50p, 100p x 100p 
	\item imaging filters: NUV, U, B, V, I
\end{enumerate}

Transfer learning is used to train the neural network models using a random selection of 80\% of the samples described in Table~\ref{tab:sample} and Table~\ref{tab:leguspubliccatalogs} separately, and the remaining 20\% are reserved for validation.
In total, this results in training samples of about 1000, 800, 600, and 1700 class 1, 2, 3, 4 objects primarily classified by BCW, and about 1400, 1800, 1800, 3900 objects with LEGUS consensus classifications.

\color{black} 

Absolute values of pixels are rescaled to be in the range \lbrack0,\,1\rbrack, to avoid the brightness of the sources from becoming a parameter in the classification. During training we use several standard data augmentation strategies, such as random flips, and random rotations in the range [$0,\, 2\pi$] to make sure that the trained neural networks are robust against those transformations. Taking into account the batch size mentioned above for \texttt{ResNet18} and \texttt{VGG19-BN}, and bearing in mind that we trained the models using about 10,000 batches, this means that the nets were exposed to 320,000 and 160,000 images, respectively. Note, however, that the data augmentation techniques used during the training stage may produce very similar images to the actual star cluster images curated for this analysis.

To investigate whether networks trained in this manner can be used to automate classification of star clusters in the PHANGS-HST dataset in the future, we test the networks on the first observations obtained by PHANGS-HST of the spiral galaxy NGC 1559.   
The PHANGS-HST NGC1559 observations provide 302, 252, 162, and 710 class 1, 2, 3, 4 objects, as classified by BCW (Table~\ref{tab:sample}).

\section{Results}
\label{sec:res}

We present four sets of results in this section. 

In Section~\ref{sec:res_1}, we present the classification accuracy for the four categories of star clusters candidates relative to classifications primarily determined by BCW and those based on the mode of classifications performed by three LEGUS team members. We also present the uncertainty quantification analysis of those models (i.e., due to random weight initialization).

In Section~\ref{sec:res_2}, we quantify the robustness of our neural network models to generalize to star cluster images in different galaxies, choosing the PHANGS-HST observations of NGC 1559 as the driver of this exercise as discussed above.

In Section~\ref{sec:res_3}, we report on whether the classification accuracy depends on the size of the images used for network training.

In Section~\ref{sec:res_4}, we report on relative importance of different filters for image classification in our resulting deep learning models.

\subsection{Does prediction accuracy depend on the origin of the classifications?}
\label{sec:res_1}

It is often useful to approach a problem using multiple methods to
check how sensitive the results are to the chosen method. For example, the use of
both ResNet18 and VGG19-BN architectures in this paper allows us to see
which one provides better results, but as we will show below, the
results are quite robust no matter which is used. We use a
similar strategy in this section by examining the results from training using two
different classification samples, namely the BCW sample (see Table 1)
and the LEGUS-consensus (3 classifiers) sample (see Table 2). While
the BCW sample might be expected to have greater internal self-consistency 
since it was performed by a single experienced classifier, averaging the results of
three less-experienced classifiers might be expected to reduce the random
noise. Hence it is not obvious which approach might give better
results in this pilot project.  In the long run, the development of a
much larger standardized database using a full range of experienced
classifiers, as discussed in Section 5, may be required to make
significant improvements.

First, we quantify the performance of our models for classification accuracy when we fine-tune the models to determine whether the transfer learning was effective at learning the morphological features that tell apart the four classes of star clusters, and to assess the robustness of the optimization procedure for image classification.  As described above, to fine-tune the models pre-trained with the \texttt{ImageNet} dataset, the weights of the last layers and the last fully connected layers of the \texttt{VGG19-BN} and \texttt{ResNet18} models are randomly initialized.  The process is performed separately for the datasets described in Tables~\ref{tab:sample} and ~\ref{tab:leguspubliccatalogs} to examine the dependence of the results on the origin of the classifications. \color{black}

\begin{figure*}
\centerline{
   \includegraphics[width=.38\linewidth]{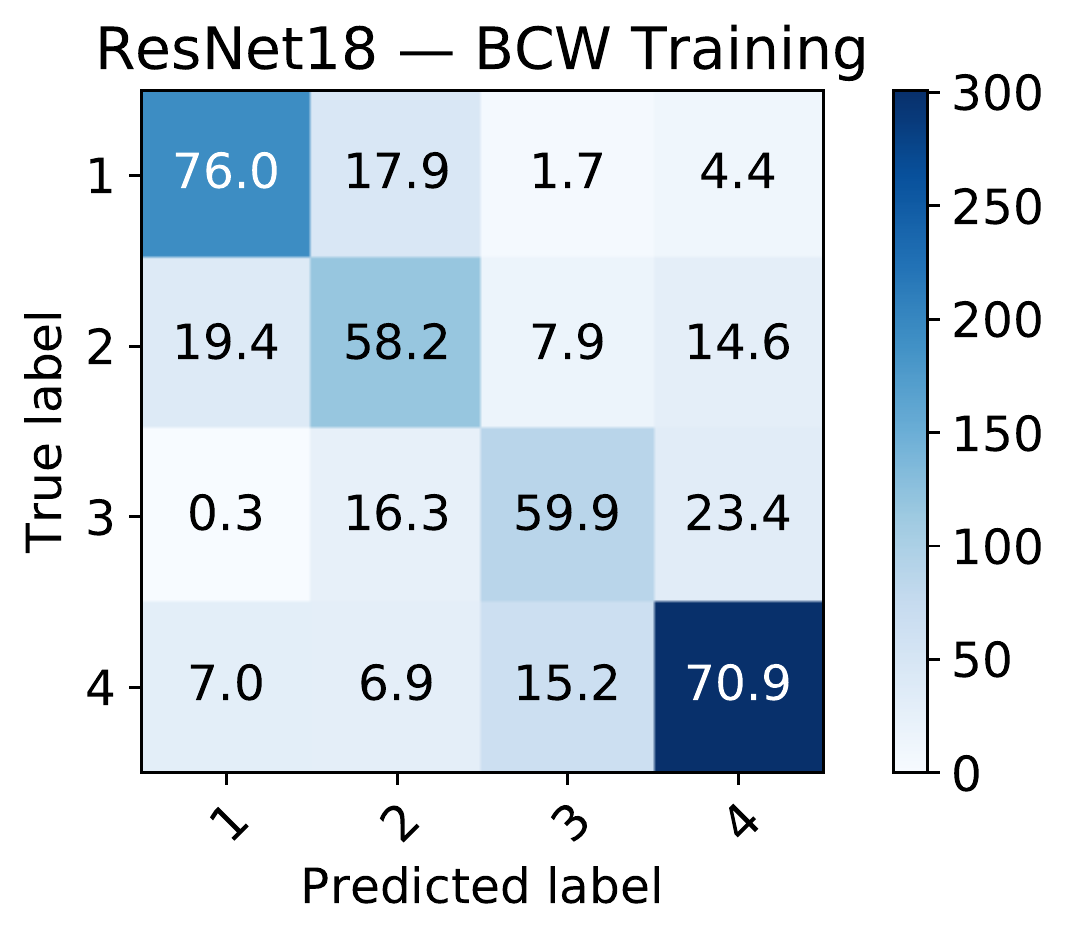}   
   \includegraphics[width=.38\linewidth]{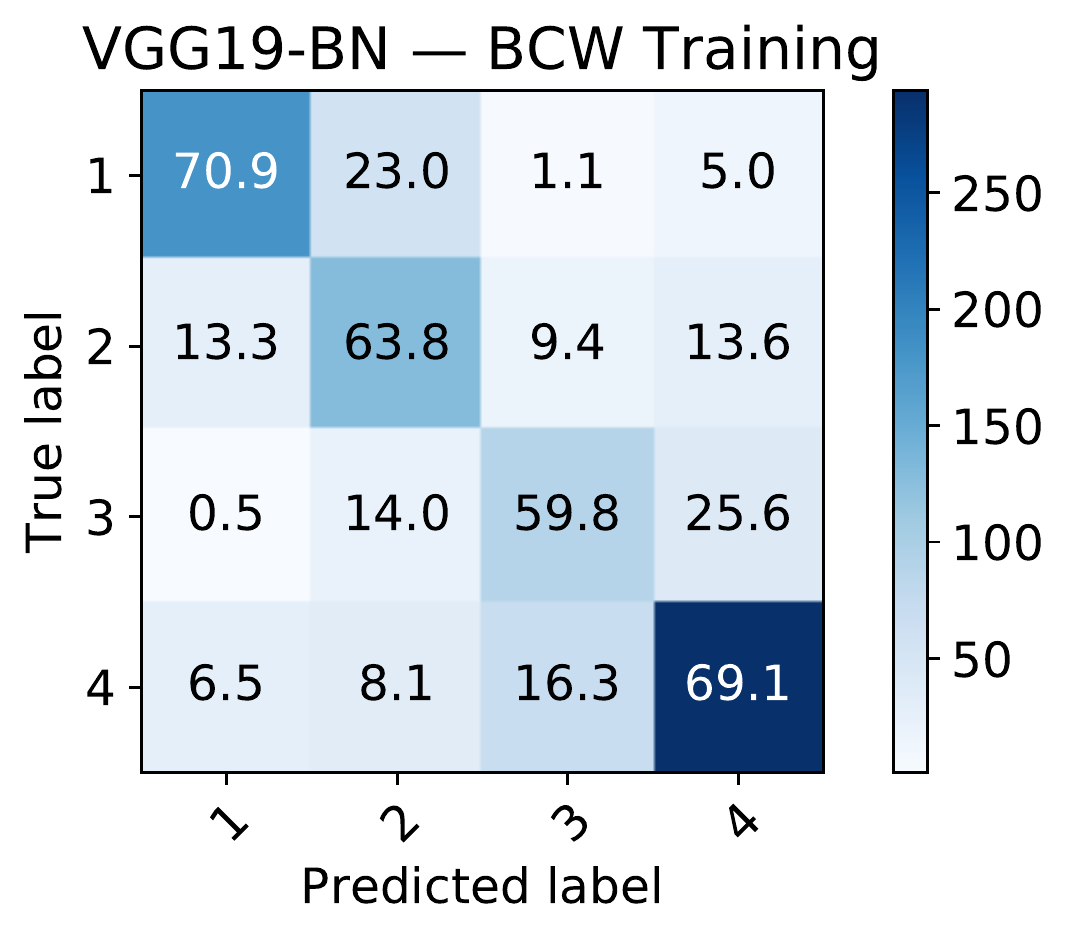} 
}
\centerline{
   \includegraphics[width=.38\linewidth]{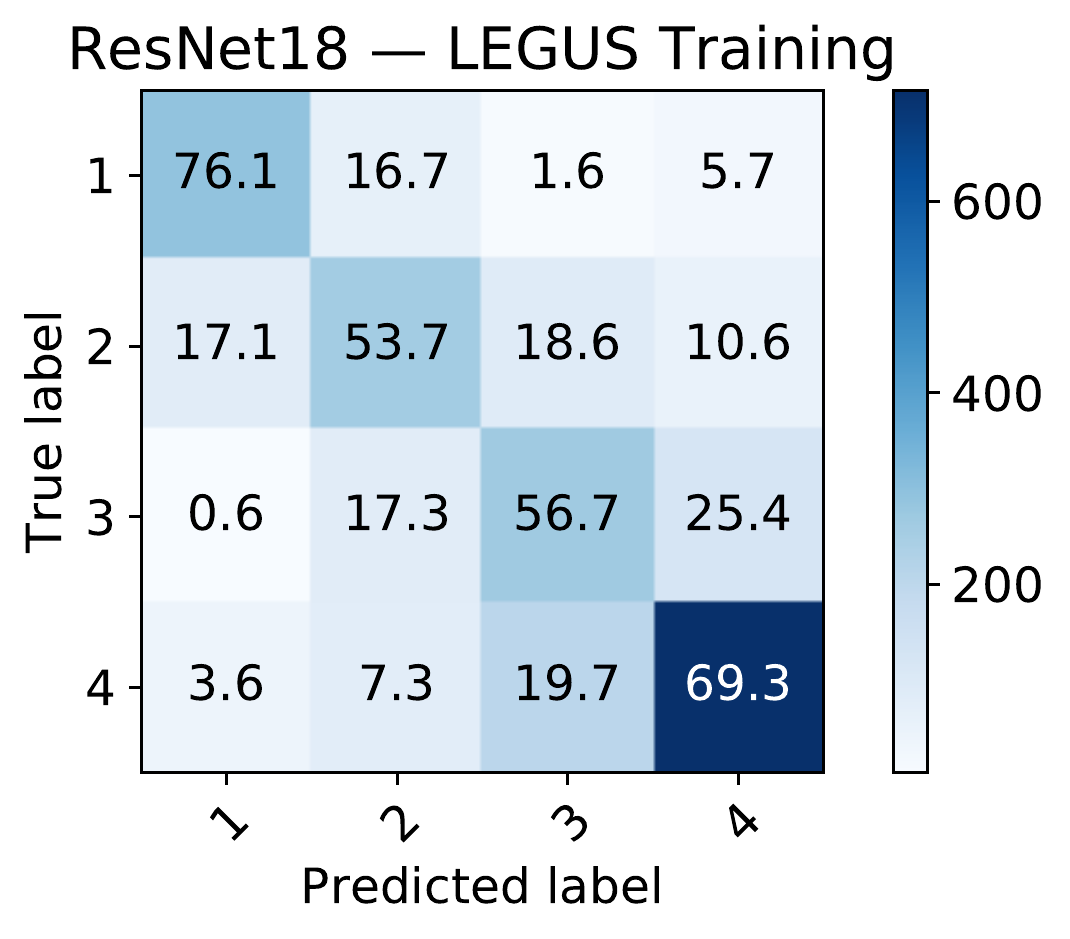}   
   \includegraphics[width=.38\linewidth]{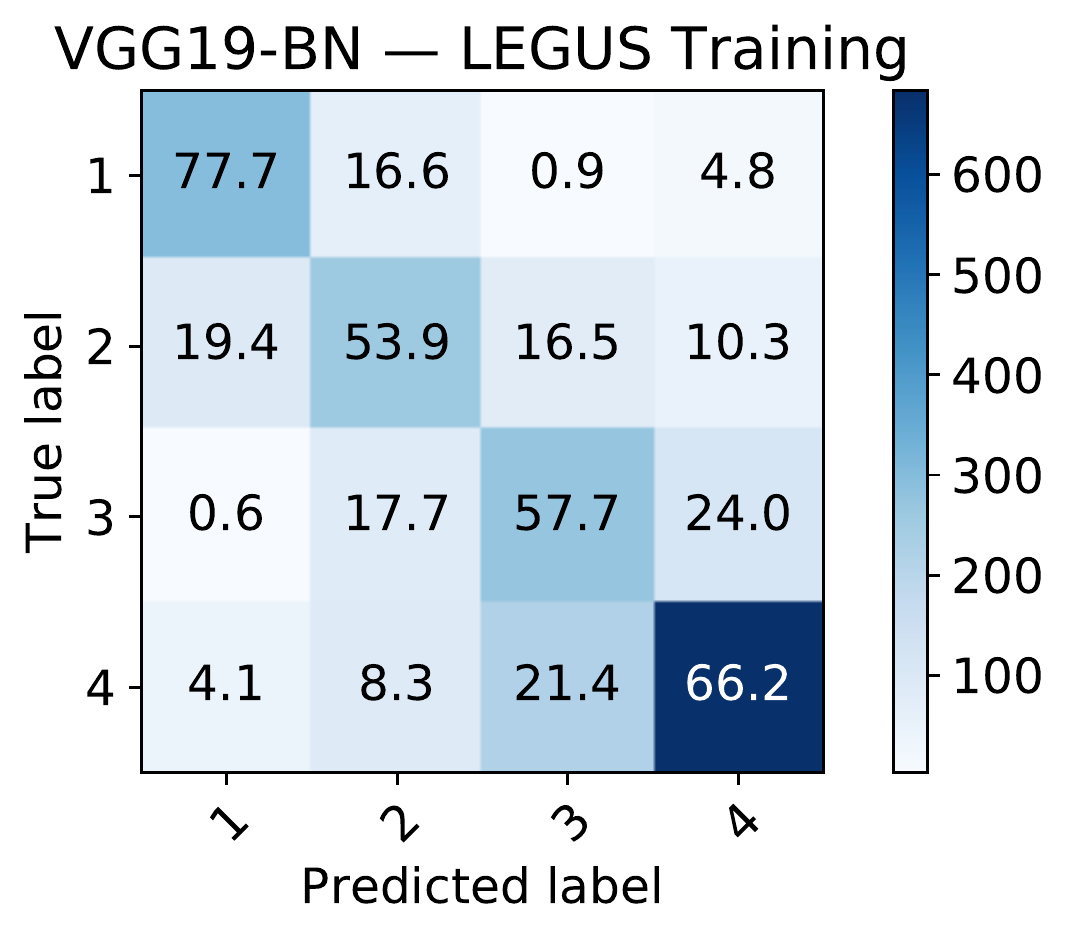} 
}
\caption{Top panels: Prediction, averaged over 10 models, of \texttt{ResNet18} (left) and \texttt{VGG19-BN} (right) trained on 80\% of the data described in Table~\ref{tab:sample} and then tested on 20\% of the data reserved for validation testing and not used for training.  Note that in these confusion matrices each column corresponds to a predicted class, whereas each row corresponds to an actual class.  Correct classification results are given along the diagonal from the top left to bottom-right of the matrices.    The color bar indicates the number of evaluation images used.  Bottom panels: Same as top panels, but for data in Table~\ref{tab:leguspubliccatalogs}.}
\label{cms}
\end{figure*}

The results based on training with classifications primarily determined by BCW are presented in the top row of confusion matrices in Figure~\ref{cms}, for both the ResNet18 and VGG19-BN models, with mean classification accuracy taken as the average over ten individual trainings from scratch.  As a reminder, the reported accuracies are based on classification of a random set of 20\% of the overall sample that was not included in the training (the "validation" sample).  Likewise, the results based on training with the mode of classifications performed by three LEGUS team members are presented in the bottom row in Figure~\ref{cms}.\color{black}

The main result is that the classification accuracies for the validation samples are comparable for both ResNet18 and VGG19-BN networks, as well as for both training samples.  Reading along the diagonal of the confusion matrices presented in Figure~\ref{cms}, for the models trained on the objects primarily classified by BCW, the accuracies for ResNet18 are 76\%, 58\%, 60\%, 71\% for class 1, 2 ,3, and 4 objects respectively, and 71\%, 64\%, 60\%, 69\% for VGG19-BN.  Similarly, for the networks trained on the mode of classifications performed by three LEGUS members the accuracies are 78\%, 54\%, 58\%, 66\% for ResNet18 and 76\%, 54\%, 57\%, 69\% for VGG19-BN.  This provides evidence that our proof-of-concept neural network models are resilient to the choice of data used for training and validation despite the fact that the two samples were (i) labelled by different classifiers; and (ii) include different parent galaxies at a wide range of distances (4-10 Mpc for the objects primarily classified by BCW, and 4-18 Mpc for the sample with LEGUS consensus classifications.) Our findings indicate that notwithstanding these seemingly important differences, the prediction accuracies using these two independent datasets are fairly consistent.  \color{black}

The variance in the ten independent classification measurements provide measure of the robustness of the models.   The variances for our neural network models trained on the classifications primarily determined by BCW are given in Tables~\ref{tab:res18_bcw} and~\ref{tab:vgg19_bcw}.  In all cases, the variances are between 4-8\%.  The variances for LEGUS classifications are comparable.\color{black}

\begin{figure*}
\centerline{
\includegraphics[width=.3\linewidth]{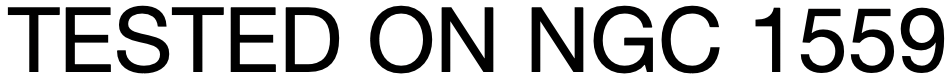}
}
\centerline{
   \includegraphics[width=.4\linewidth]{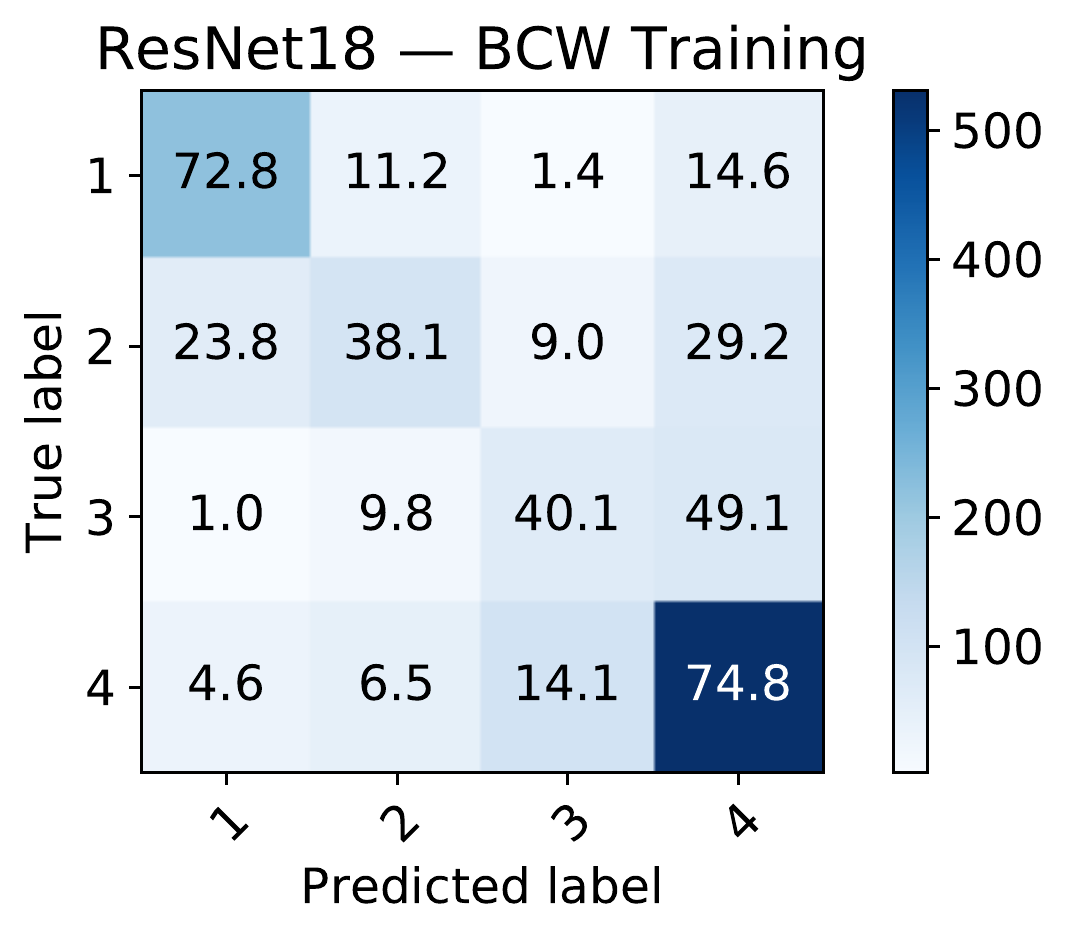}   
   \includegraphics[width=.4\linewidth]{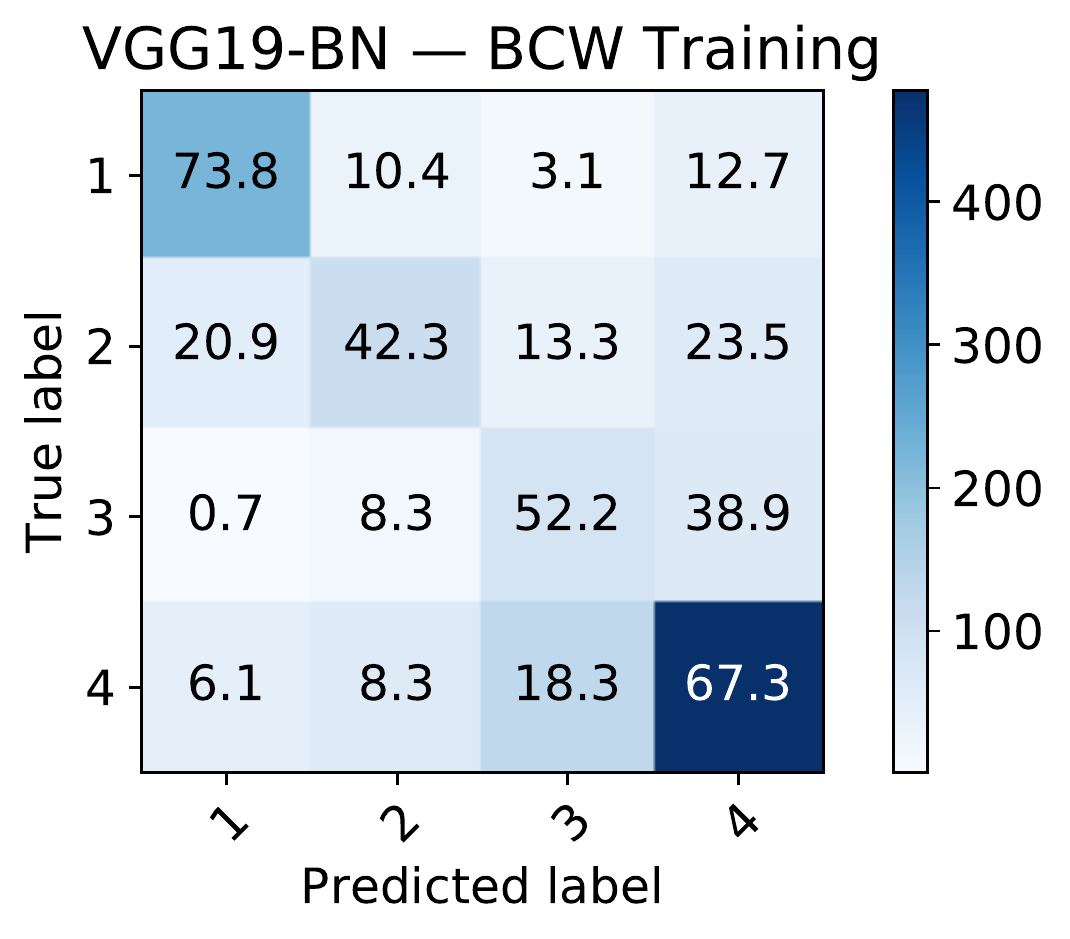} 
}
\centerline{
    \includegraphics[width=.4\linewidth]{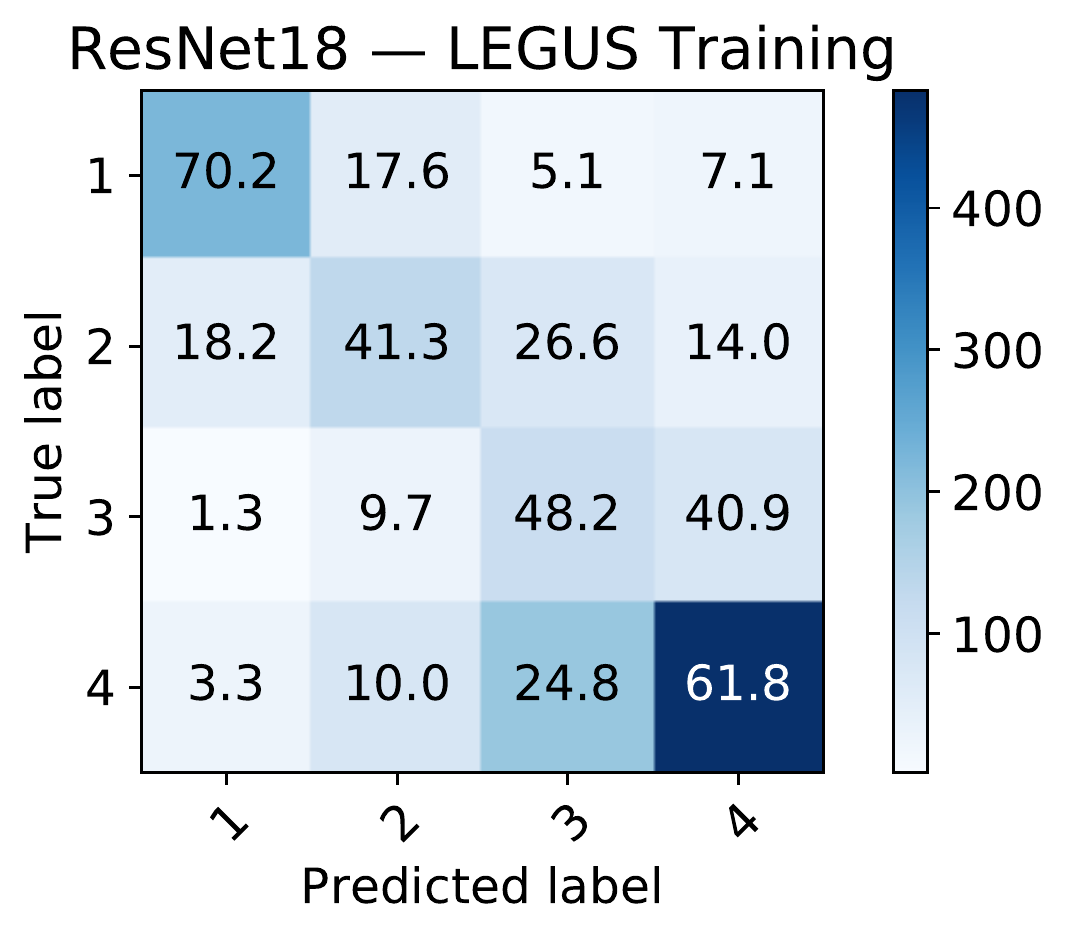}
    \includegraphics[width=.4\linewidth]{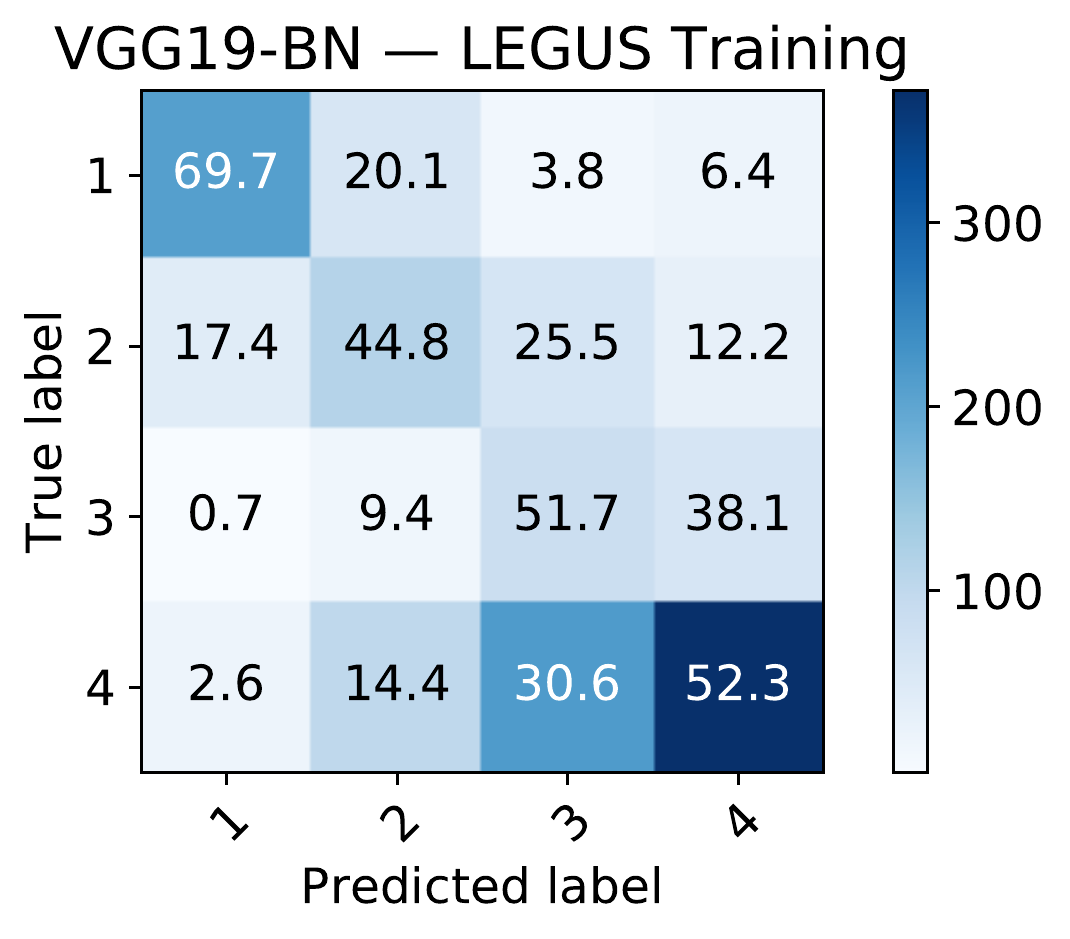}   
 }
\caption{Top panels: Same as Figure~\ref{cms}, but now the models trained on the classifications primarily determined by BCW (Table~\ref{tab:sample}) are applied to predict classifications for candidates in PHANGS-HST observations of NGC 1559, a galaxy which was not included in the training samples.  As before, results were obtained after averaging over 10 models. Bottom panels: Same as top row, but for models trained on the mode of classifications performed by three LEGUS team members (Table~\ref{tab:leguspubliccatalogs}).}
\label{fig:ngc1559}
\end{figure*}

\begin{table*}
\begin{tabular}{llllll}
\hline
\hline
                 &           \multicolumn{1}{c}{Class 1 [\%]} & \multicolumn{1}{c}{Class 2 [\%]} & \multicolumn{1}{c}{Class 3 [\%]} & \multicolumn{1}{c}{Class 4 [\%]} & \multicolumn{1}{c}{\textbf{Total}}\\
\hline

BCW Class 1          & \textbf{76.0}$\pm$ 4.2       & 17.9$\pm$ 4.4      & 1.7$\pm$0.7      & 4.4$\pm$1.4 &    254  \\
BCW Class 2          & 19.4 $\pm$3.5     & \textbf{58.2}$\pm$5.3     & 7.9$\pm$3.5     & 14.6$\pm$3.0      & 202      \\
BCW Class 3          & 0.3 $\pm$0.5      & 16.3$\pm$5.4     & \textbf{59.9}$\pm$6.8     & 23.4$\pm$5.6      & 147       \\
BCW Class 4         & 7.0$\pm$2.1      & 6.9$\pm$2.9       & 15.2$\pm$3.1      & \textbf{70.9}$\pm$4.8     & 425        \\
\end{tabular}
\caption{Prediction of \texttt{ResNet18} on  20\% of the data in Table~\ref{tab:sample} reserved for validation testing and not included in the training,  averaged over 10 models. The averaged predictions from Figure~\ref{cms} are repeated, but now the standard deviations are also shown.  The number of validation images for each class are listed in the final column.  \color{black}} 
\label{tab:res18_bcw}
\end{table*}


\begin{table*}
\begin{tabular}{llllll}
\hline
\hline
                 &           \multicolumn{1}{c}{Class 1 [\%]} & \multicolumn{1}{c}{Class 2 [\%]} & \multicolumn{1}{c}{Class 3 [\%]} & \multicolumn{1}{c}{Class 4 [\%]} & \multicolumn{1}{c}{\textbf{Total}}\\
\hline

BCW Class 1          & \textbf{70.9}$\pm$6.2 & 23.0$\pm$4.8 &  1.1$\pm$0.7 & 5.0$\pm$1.9  &    254  \\
BCW Class 2          &   13.3$\pm$4.3 & \textbf{63.8}$\pm$4.8 &  9.4$\pm$2.9 & 13.6$\pm$3.6   & 202      \\
BCW Class 3          &   0.5$\pm$0.7 & 14.0$\pm$6.3 & \textbf{59.8}$\pm$7.5 & 25.6$\pm$7.4   & 147       \\
BCW Class 4         &   6.5$\pm$2.4 &  8.1$\pm$2.6 & 16.3$\pm$3.8 & \textbf{69.1}$\pm$6.8   & 425        \\
\end{tabular}
\caption{As Table~\ref{tab:res18_bcw}, but now using \texttt{VGG19-BN}.}
\label{tab:vgg19_bcw}
\end{table*}

\begin{table*}
\begin{tabular}{llllll}
\hline
\hline
                 &           \multicolumn{1}{c}{Class 1 [\%]} & \multicolumn{1}{c}{Class 2 [\%]} & \multicolumn{1}{c}{Class 3 [\%]} & \multicolumn{1}{c}{Class 4 [\%]} & \multicolumn{1}{c}{\textbf{Total}}\\
\hline

BCW Class 1          & \textbf{72.8}$\pm$ 7.6       & 11.2$\pm$ 3.8      & 1.4$\pm$0.6      & 14.6$\pm 5.1$ &    302  \\
BCW Class 2          & 23.8$\pm$4.3     & \textbf{38.1}$\pm$5.9     & 9.0$\pm$4.0     & 29.2$\pm$4.7      & 252      \\
BCW Class 3          & 1.0$\pm$0.5      & 9.8$\pm$4.2     & \textbf{40.1}$\pm$7.1     & 49.1$\pm$6.1      & 162       \\
BCW Class 4         & 4.6$\pm$1.4      & 6.5$\pm$1.8       & 14.1$\pm$3.1      & \textbf{74.8}$\pm$3.5     & 710        \\
\end{tabular}
\caption{Prediction of \texttt{ResNet18} trained on star clusters primarily classified by BCW (Table~\ref{tab:sample}) for candidates in spiral galaxy NGC 1559 from the PHANGS-HST program, averaged over 10 models. Each row shows the averaged predictions (same as shown in top left panel of Figure~\ref{fig:ngc1559}), but now together with the standard deviations from the 10 models.  The numbers of objects classified are given in the last column. This experiment was performed to test the ability of this neural network model to generalize to images from galaxies not included in the training sample.  It is notable that NGC 1559 is roughly twice as far away as any of galaxies in the BCW training sample.} 
\label{tab:res18_ngc1559}
\end{table*}

\begin{table*}
\begin{tabular}{llllll}
\hline
\hline
                 &           \multicolumn{1}{c}{Class 1 [\%]} & \multicolumn{1}{c}{Class 2 [\%]} & \multicolumn{1}{c}{Class 3 [\%]} & \multicolumn{1}{c}{Class 4 [\%]} & \multicolumn{1}{c}{\textbf{Total}}\\
\hline

BCW Class 1     &     \textbf{73.8}$\pm$4.8 & 10.4$\pm$3.5 & 3.1$\pm$1.3 & 12.7$\pm$4.4 &    302  \\
BCW Class 2          &   20.9$\pm$6.4 & \textbf{42.3}$\pm$7.9 & 13.3$\pm$2.6 & 23.5$\pm$8.0    & 252      \\
BCW Class 3          &    0.7$\pm$0.6 & 8.3$\pm$3.3 & \textbf{52.2}$\pm$5.9 & 38.9$\pm$7.5   & 162       \\
BCW Class 4         & 6.1$\pm$2.4 &  8.3$\pm$3.3 & 18.3$\pm$3.0 & \textbf{67.3}$\pm$6.8    & 710        \\
\end{tabular}
\caption{As Table~\ref{tab:res18_ngc1559}, but now using \texttt{VGG19-BN}. }
\label{tab:vgg19_ngc1559}
\end{table*}

\subsection{How accurately can the models predict classifications for clusters in galaxies not included in the training sample?}
\label{sec:res_2}

To further assess the robustness and resilience of our neural network models, we use them to classify images from a galaxy not included in the original training dataset, namely the PHANGS-HST target NGC 1559. This galaxy is about two to four times further away than the galaxies in either of the training samples, with the notable exception of NGC1566, which is at a comparable distance to NGC 1559 (18 Mpc vs. 19 Mpc), and included the sample with consensus classifications from three LEGUS team members (Table~\ref{tab:leguspubliccatalogs}).  Results are presented in Figure~\ref{fig:ngc1559} and  Tables~\ref{tab:res18_ngc1559} and~\ref{tab:vgg19_ngc1559}.

Notwithstanding these differences, we again notice that all models produce comparable results.  Reading along the diagonal of the confusion matrices presented in Figure~\ref{fig:ngc1559}, for the models trained on the objects primarily classified by BCW, the accuracies for ResNet18 are 73\%, 38\%, 40\%, 75\% for class 1, 2 ,3, and 4 objects respectively, and 74\%, 42\%, 52\%, 67\% for VGG19-BN.  Likewise, for the networks trained on the mode of classifications performed by three LEGUS members the accuracies are 70\%, 41\%, 48\%, 62\% for ResNet18 and 70\%, 45\%, 52\%, 52\% for VGG19-BN. 

For all models, the performance for NGC1559 class 1 star clusters is at or above the 70\% level. The classification accuracy of the BCW-based models is similar to their performance on the validation samples (i.e., Figure~\ref{cms}).  Meanwhile for NGC1559 class 1 star clusters the performance of the models trained on the LEGUS consensus classifications are 6-8\% lower relative to the classification of the validation samples.  On the other hand for class 2 star clusters, the accuracies hover around the 40\% level, and are the lowest of the four classes.  The accuracies for the models trained on the objects primarily classified by BCW drop by $\sim$20\%: from 58\% (test subset sample) to 38\% (NGC 1559) for ResNet18, and from 64\% to 42\% for VGG19-BN.  Similarly, those trained on the LEGUS consensus classifications drop, although by only $\sim$10\%: from 54\% to 41\% for ResNet18, and from 54\% to 45\% for VGG19-BN.   The accuracies for the NGC 1559 class 3 star clusters are at the 40-50\% level, a $\sim$10\% drop for all models relative to the performance on the test subsets.  Finally, for the class 4 non-clusters, the models trained on the objects primarily classified by BCW perform comparably, i.e., at the 70\% level, while those trained on the LEGUS consensus classifications drop to the 50-60\% level.

\color{black}

\subsubsection{Uncertainty calculations through entropy analysis }
\begin{figure}
    \centering
    \includegraphics[width=\linewidth]{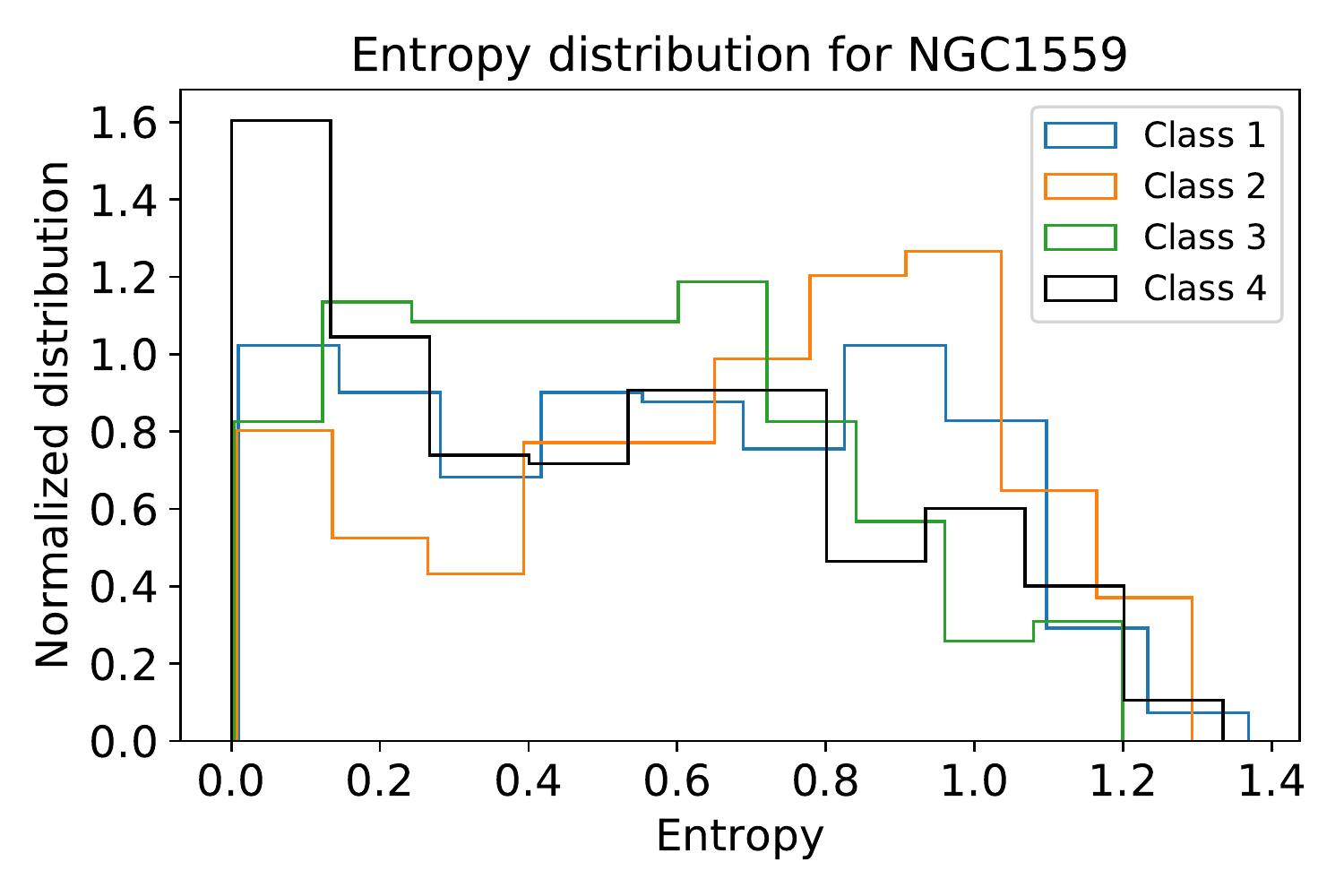}
    \caption{The uncertainty in the neural network's prediction is quantified by the entropy of the predicted probability 
    distribution over the 4 star cluster image classes considered in this analysis. For a random guess over the 4 classes, the entropy is $\ln(4)\approx 1.39$. The lower the entropy, the higher the confidence the neural network has about its prediction. The panel shows the predicted entropy value for each NGC1559 image with which we classified with our \texttt{VGG19-BN} model, trained on the objects primarily classified by BCW given in Table~\ref{tab:sample}. The x-axis shows the binned values of the entropy values, whose frequency of occurrence is indicated on the y-axis. To make clear that the area of each histogram is normalized to one, the y-axis label is explicitly labeled ``Normalized distribution."}
    \label{fig:entropy_dist}
\end{figure}

Another method to investigate the uncertainty in the models' predictions is through the computation of entropy by using the probability distributions for each of the cluster classes we are trying to classify, which is an output of the models.  Intuitively, the more pronounced the peak is in the probability distribution, the more confident the neural network is about its prediction, and in this case, the entropy calculated from the prediction probability distribution will be lower. For example, if the probability distribution is only concentrated on one class, the network network in this case is $100\%$ certain about its prediction and the entropy would be zero, i.e., there is no uncertainty. On the other hand, if the prediction assigned the same probability for all the 4 classes under consideration equally, we would have maximum uncertainty in this case, since for the given input image, all the 4 classes are equally possible to be the predicted classes, and in this case, the maximum entropy is  $\ln(4)\approx 1.39$. Figure~\ref{fig:entropy_dist} shows the distribution of the entropies for the predictions of \texttt{VGG19-BN} when tested on NGC 1559 images.

\subsection{How does classification accuracy depend on size of training images?}
\label{sec:res_3}

\begin{figure*}
\centerline{
    \includegraphics[width=.38\linewidth]{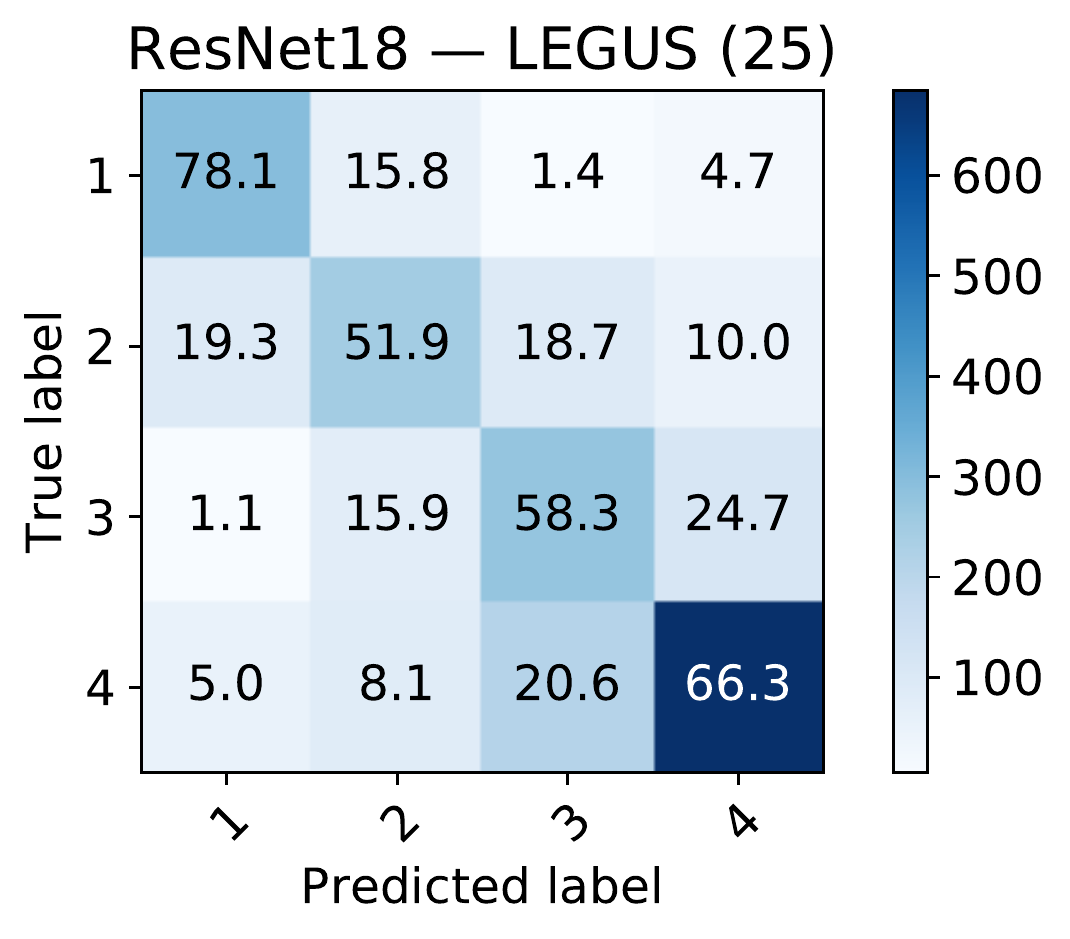}
    \includegraphics[width=.38\linewidth]{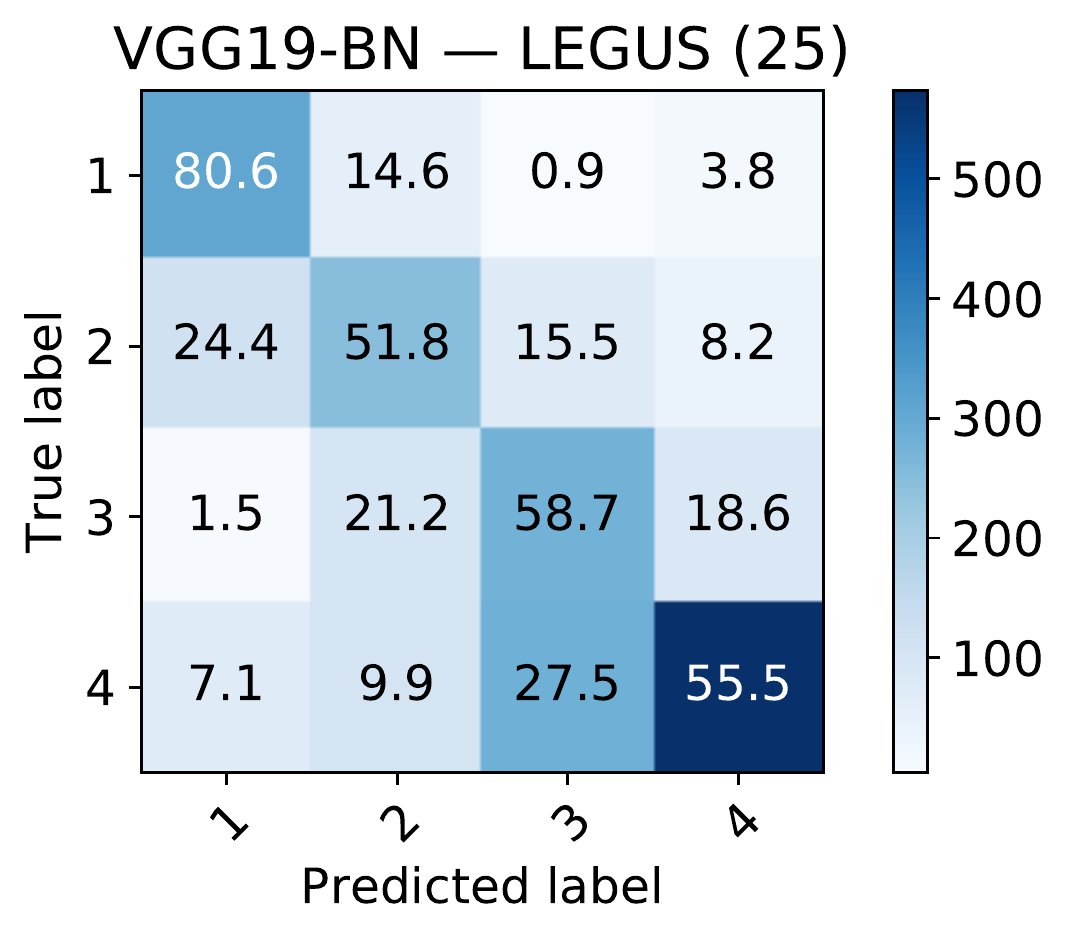}   
 }
 \centerline{
    \includegraphics[width=.38\linewidth]{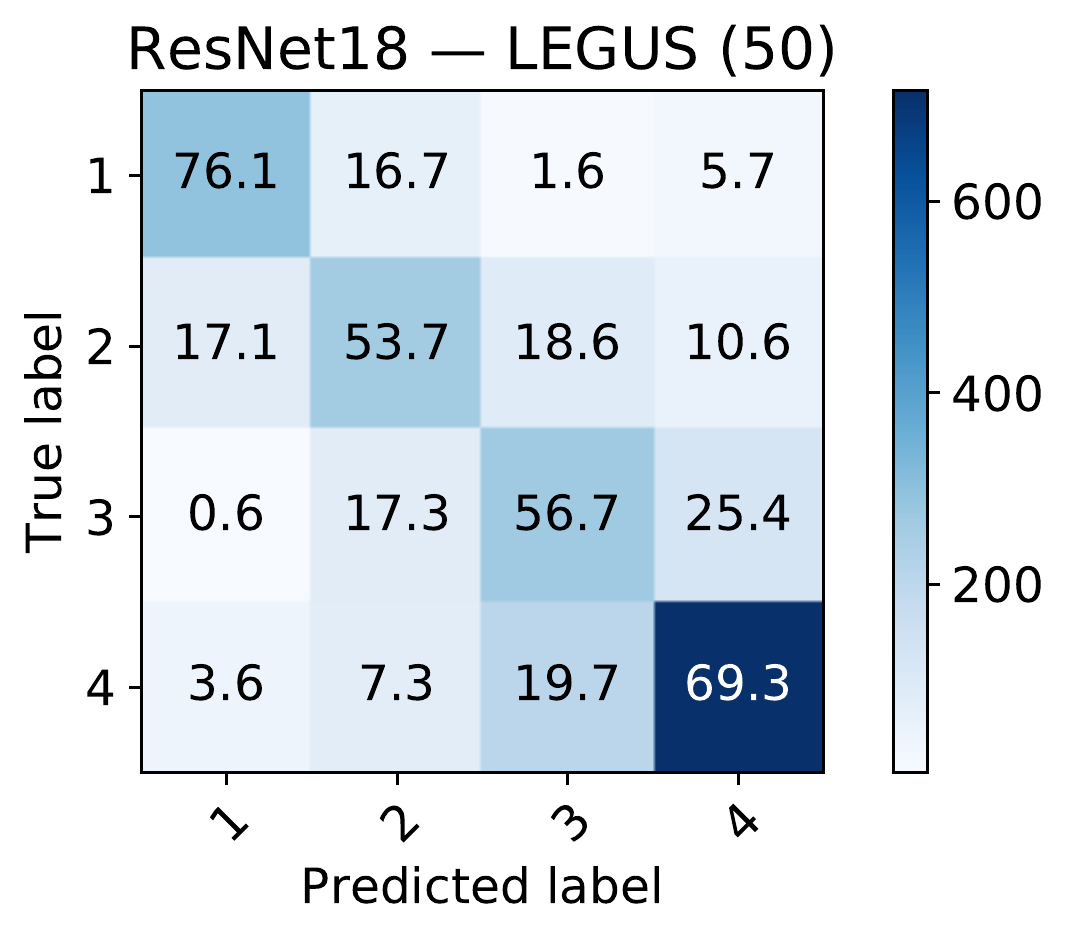}
    \includegraphics[width=.38\linewidth]{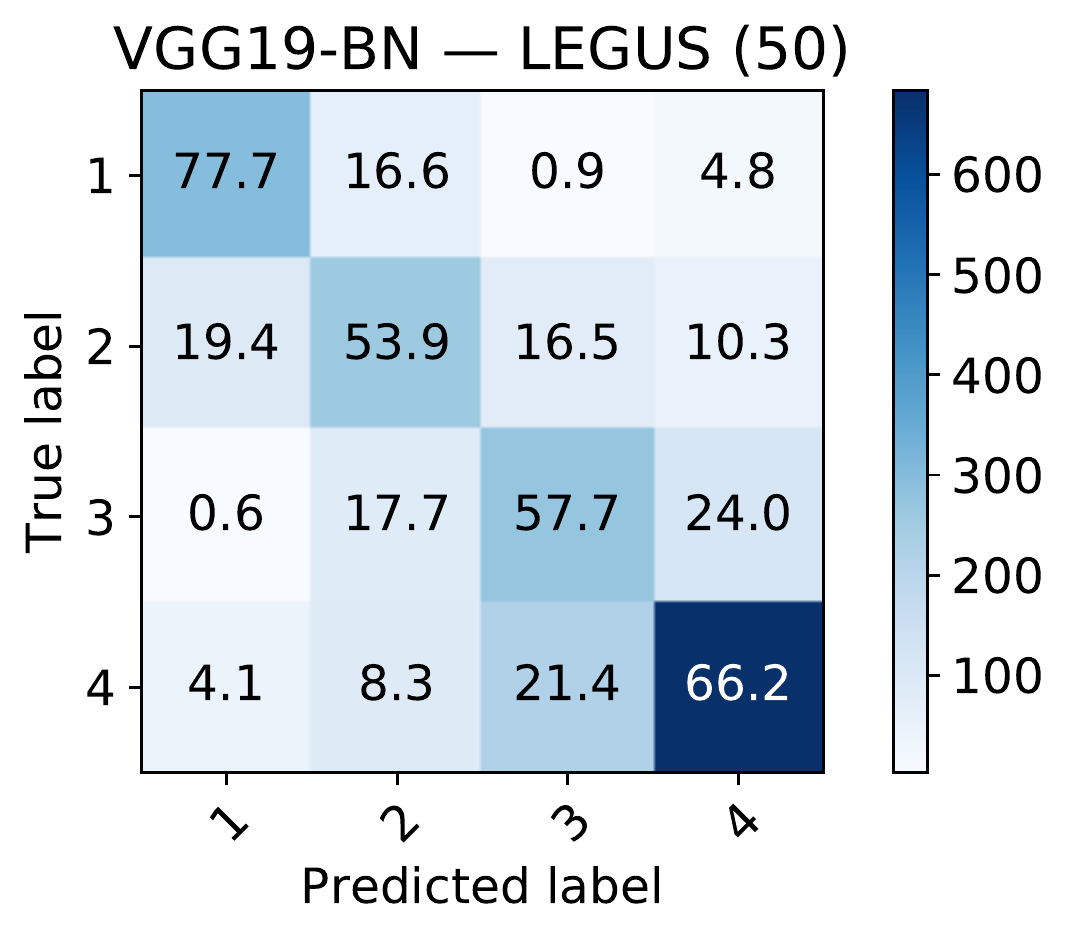}   
 }
 \centerline{
    \includegraphics[width=.38\linewidth]{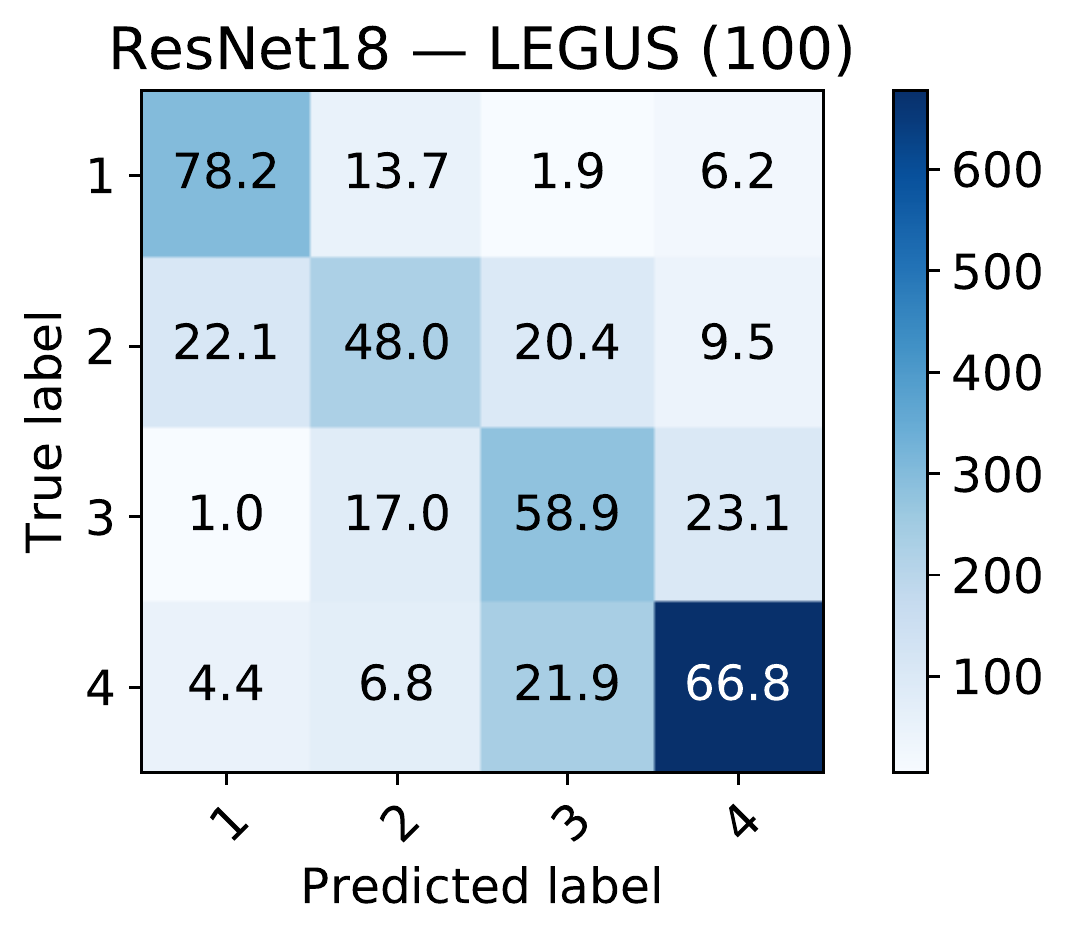}
    \includegraphics[width=.38\linewidth]{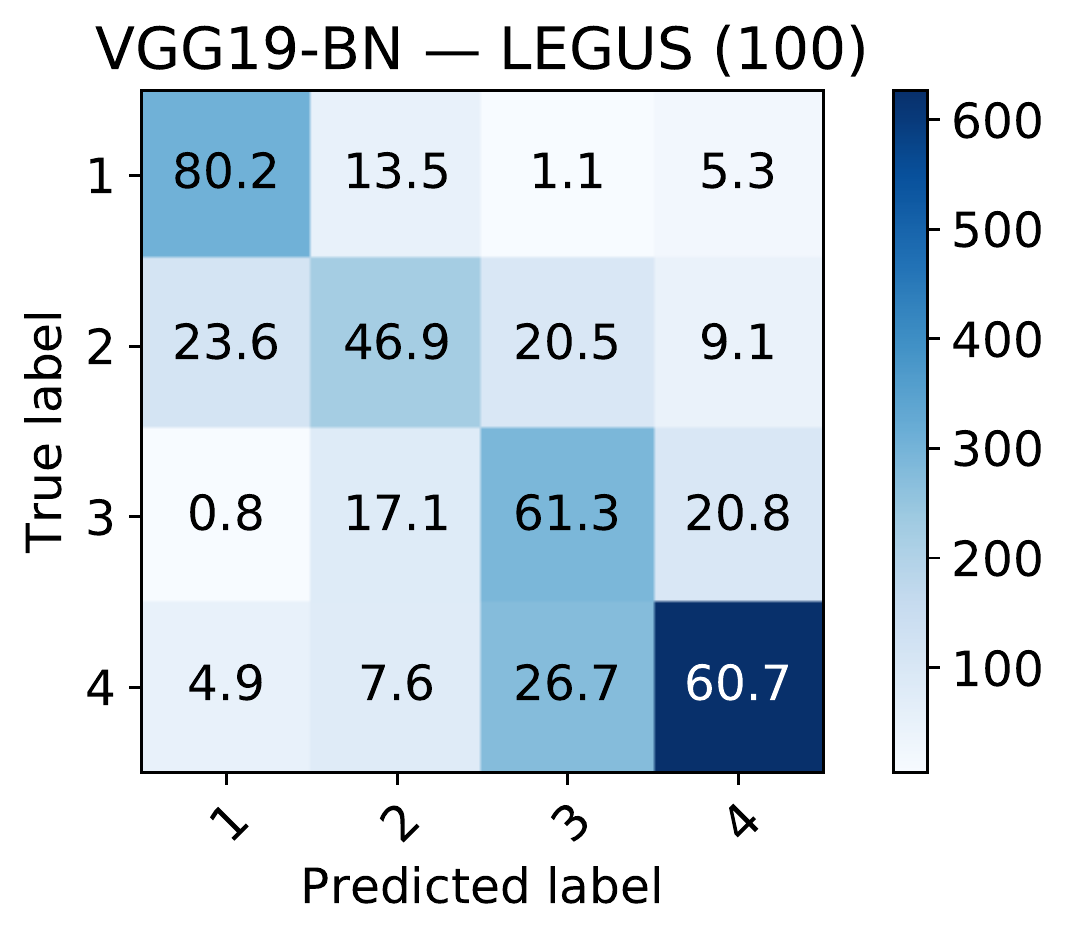}   
 }
 
\caption{ Left column: \texttt{VGG19-BN} model classification results for cropping size $25\times 25$, $50\times 50$ and $100\times 100$. Right column: as before, but now for \texttt{ResNet}.\color{black}}
\label{fig:other_LEGUS_crop_size}
\end{figure*}

To quantify the importance of image size for star cluster classification, we train our neural network models again, but with two additional cropping sizes: \(25\times25\) pixels and \(100\times100\) pixels.  In Figure~\ref{fig:other_LEGUS_crop_size}, we present results from training on the sample with LEGUS consensus classifications (again, where 80\% of the sample is used for training and 20\% for testing), where the results presented earlier from our fiducial experiments with \(50\times50\) pixels postage stamps are repeated to facilitate comparison.  We present results based on the LEGUS consensus classifications as the range of distances of the galaxies (from 3.1 Mpc to 18 Mpc; Table~\ref{tab:leguspubliccatalogs}) is inclusive of the range spanned by the sample primarily classified by BCW (Table~\ref{tab:sample}).  Hence, the physical scales subtended by the cropped images span from 16 pc (for \(25\times25\) pixel images at 3.1 Mpc) to 360 pc (for \(100\times100\) pixel images at 18 Mpc).

There are no significant differences between the results for the different cropping sizes.  These results indicate that our neural network models are resilient to this particular data curation choice.  We see variations at the level of \(\sim5\%\), which is within the expected variation in the performance of the neural network models due to random weight initialization, as indicated in Tables~\ref{tab:res18_bcw} and~\ref{tab:vgg19_bcw}.  Results for the models trained with objects primarily classified by BCW are consistent.  The results also do not change if the neural network models are trained with postage stamps using {\it random} cropping sizes ranging from \(25\times25\) pixels to \(100\times100\) pixels (i.e., a random cropping size is chosen for each object in the training and testing sample). \color{black}

\subsection{Classification accuracy as a function of imaging filter}
\label{sec:res_4}

\begin{figure*}
\centerline{
    \includegraphics[width=.38\linewidth]{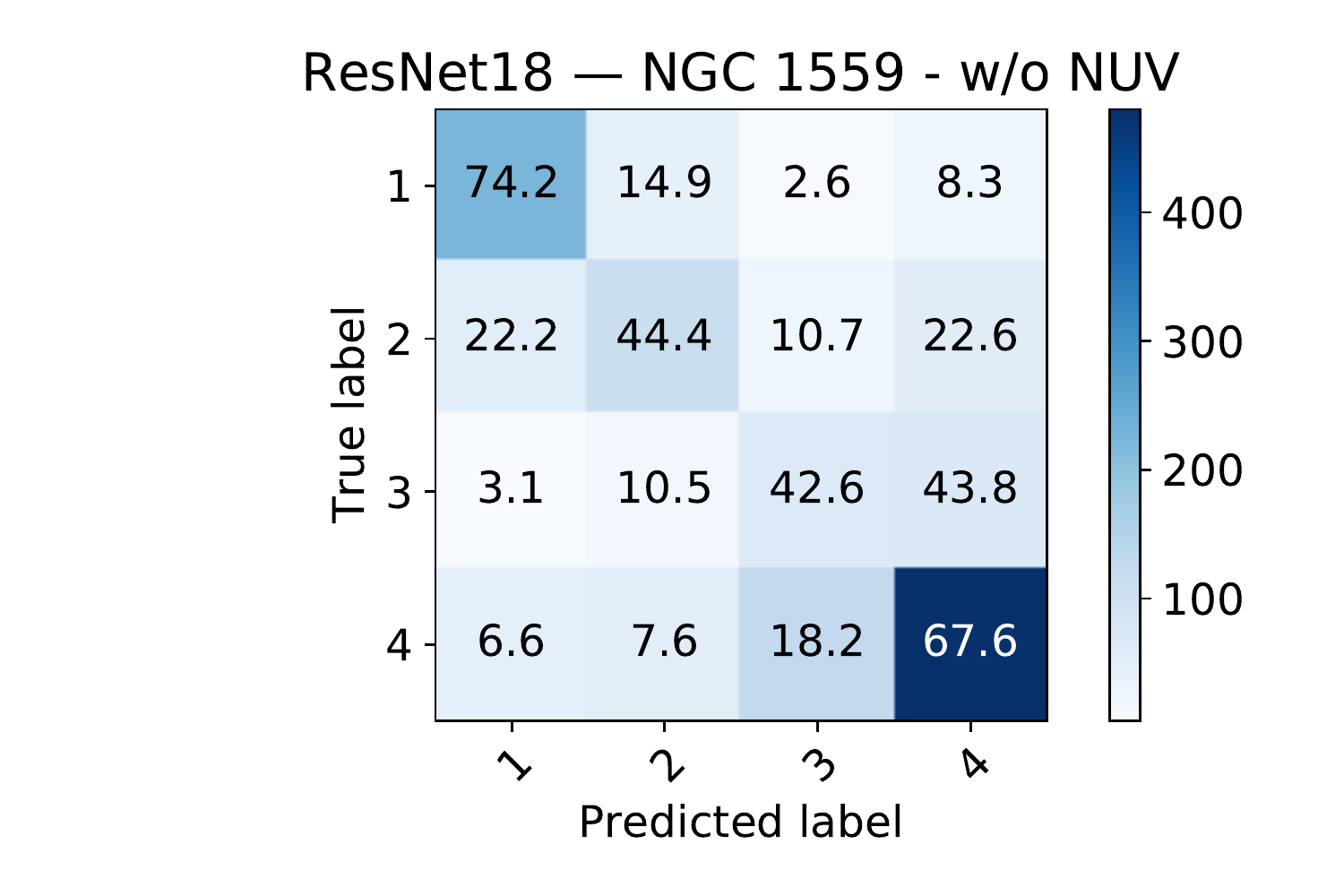}   
    \includegraphics[width=.38\linewidth]{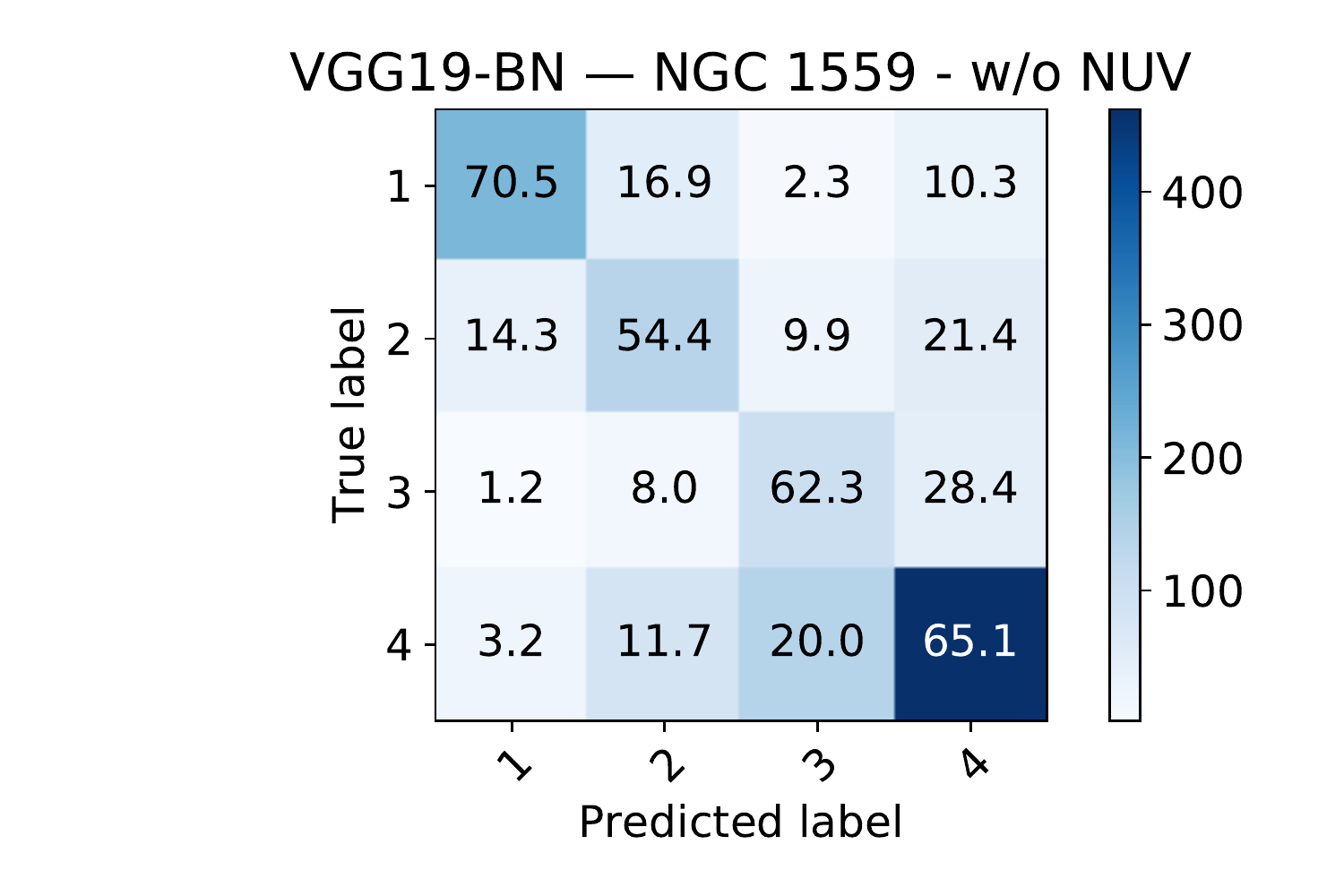}
 }
 \centerline{
    \includegraphics[width=.38\linewidth]{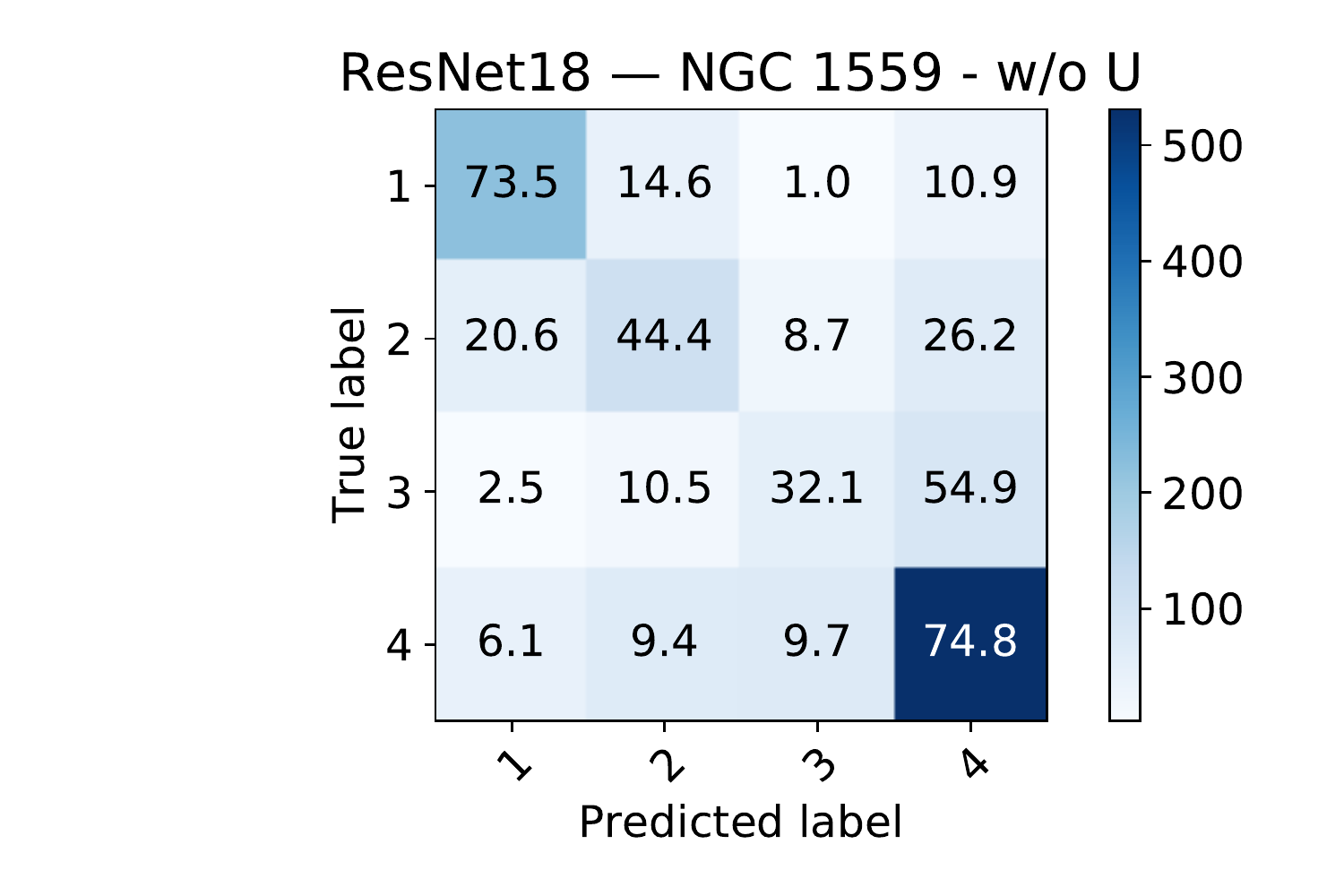}   
    \includegraphics[width=.38\linewidth]{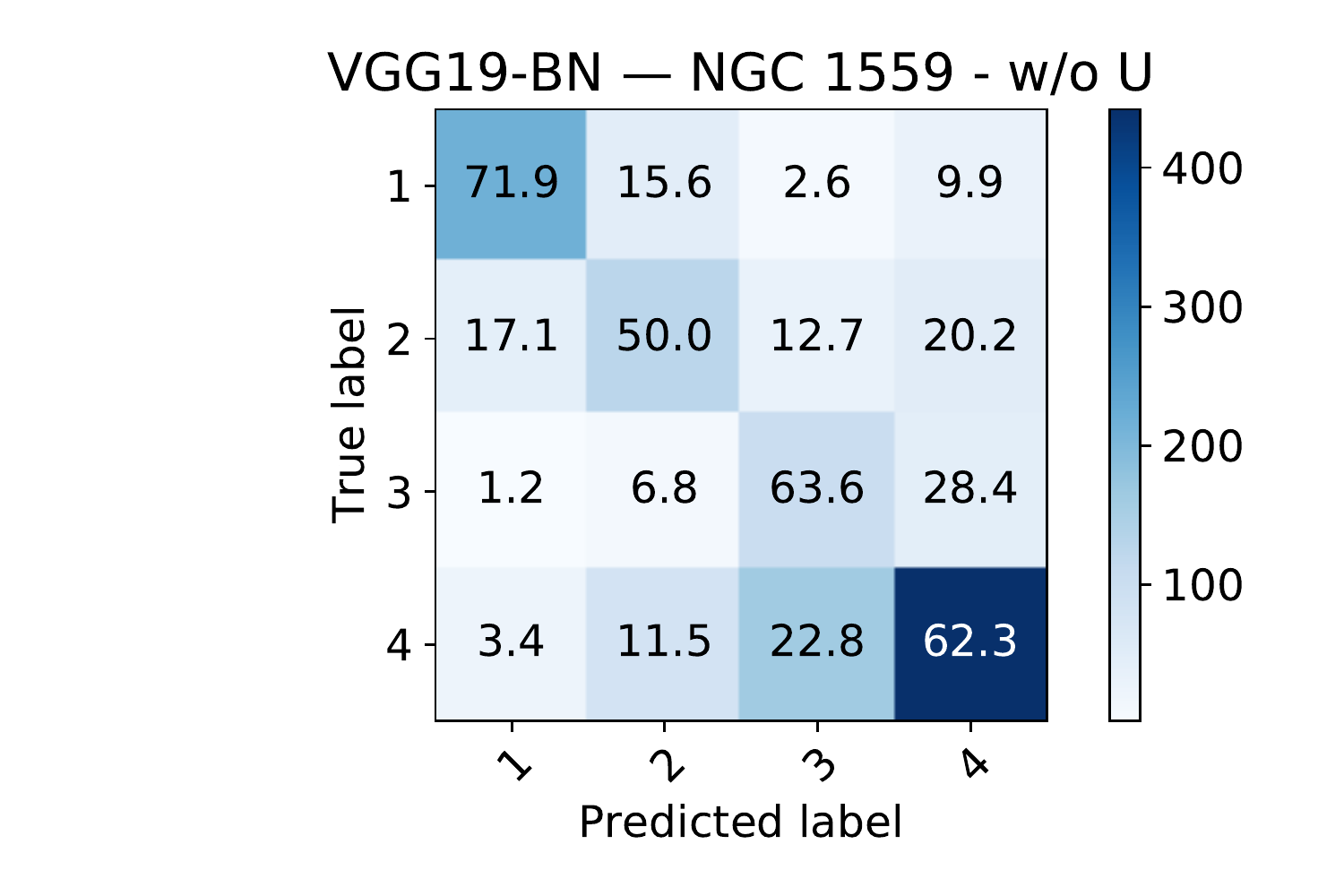}
 }
 \centerline{
    \includegraphics[width=.38\linewidth]{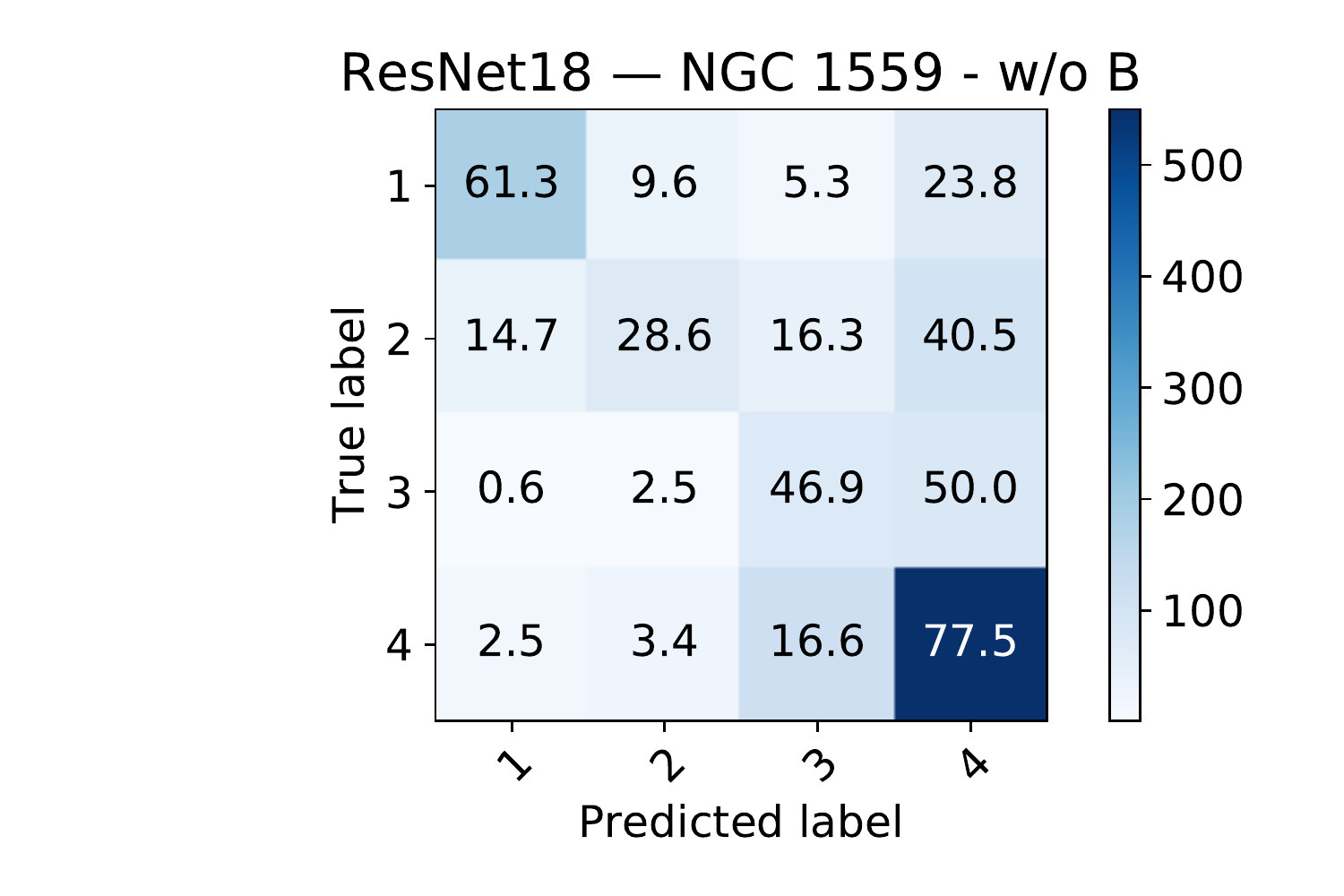}   
    \includegraphics[width=.38\linewidth]{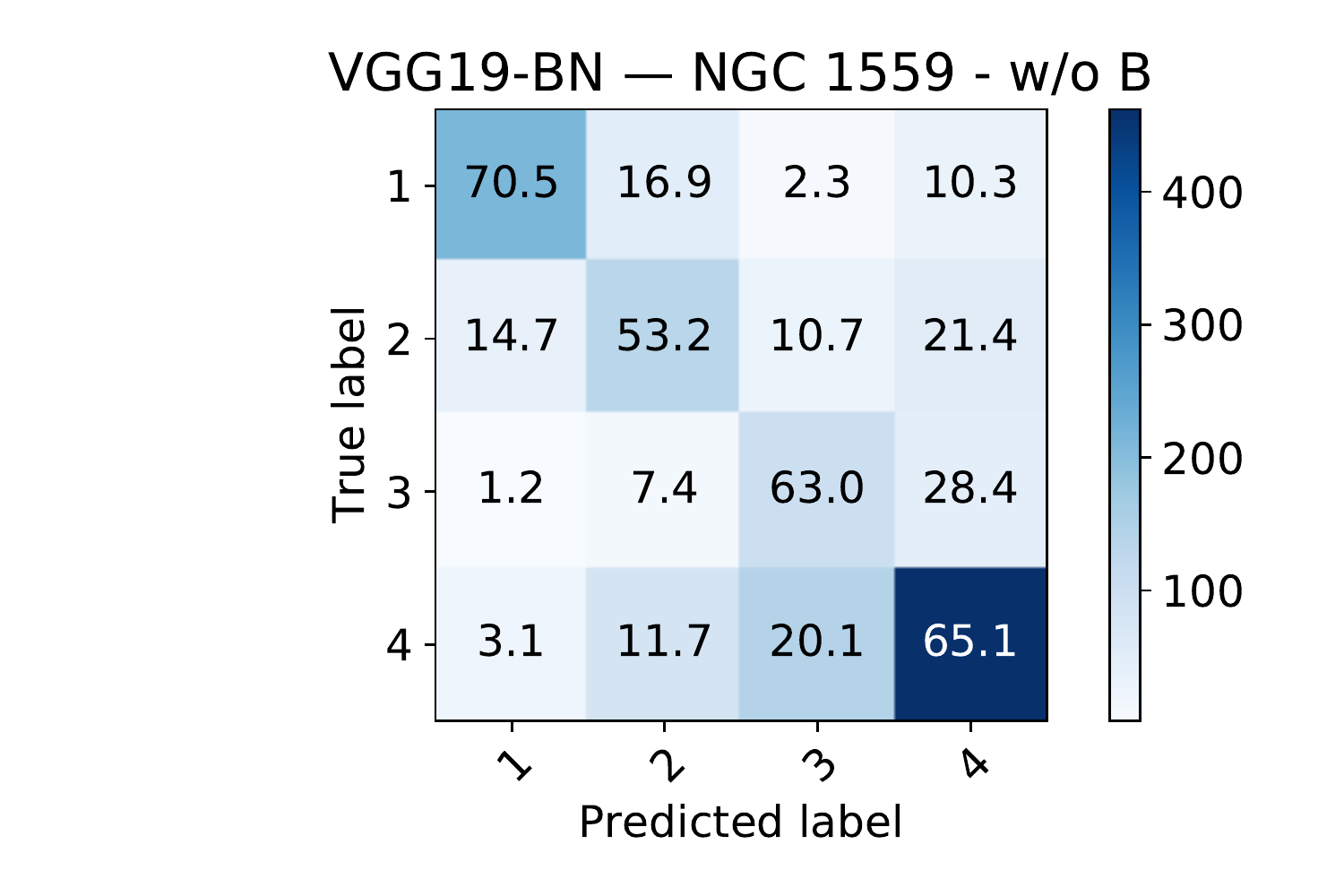}
 }
  \centerline{
    \includegraphics[width=.38\linewidth]{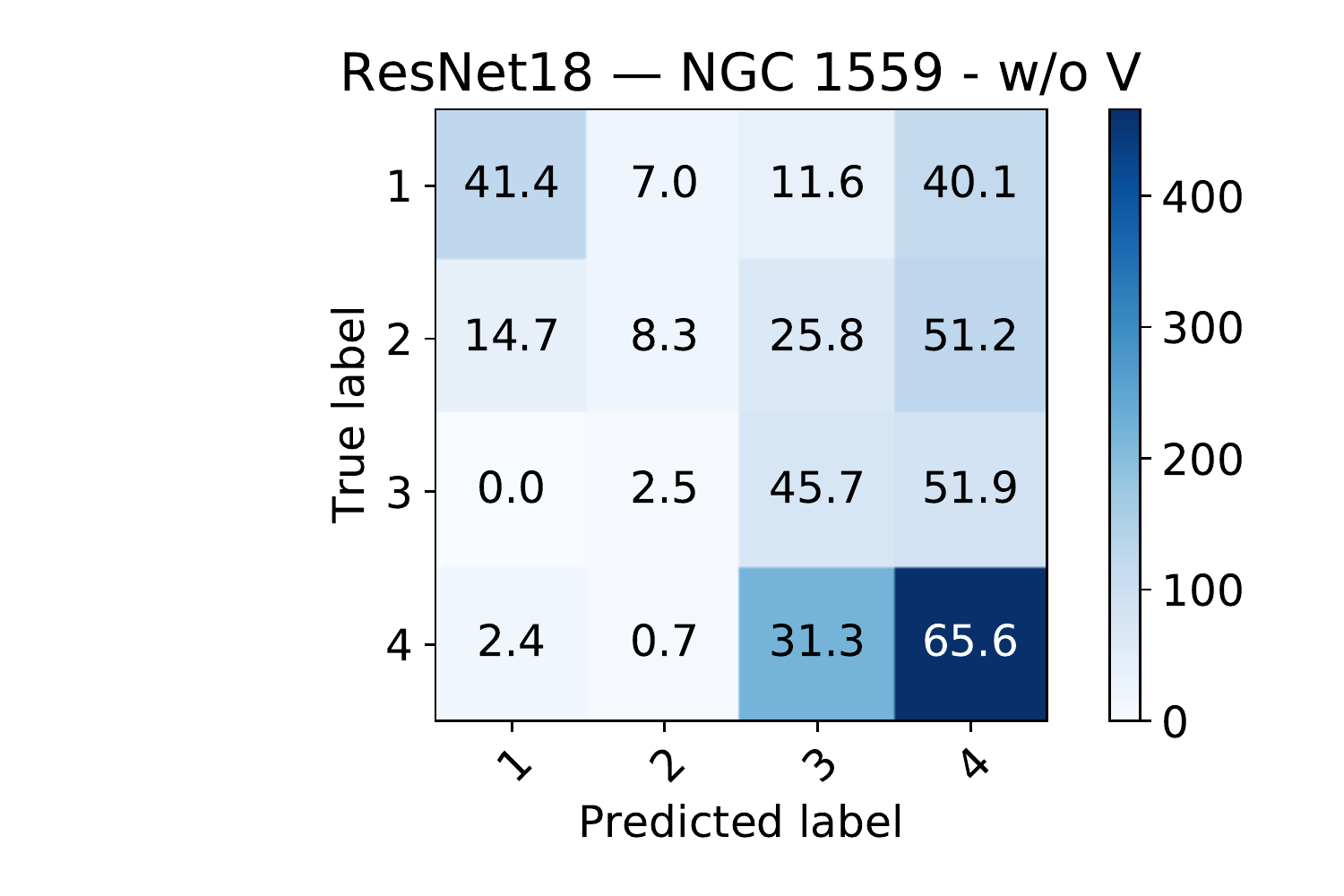}   
    \includegraphics[width=.38\linewidth]{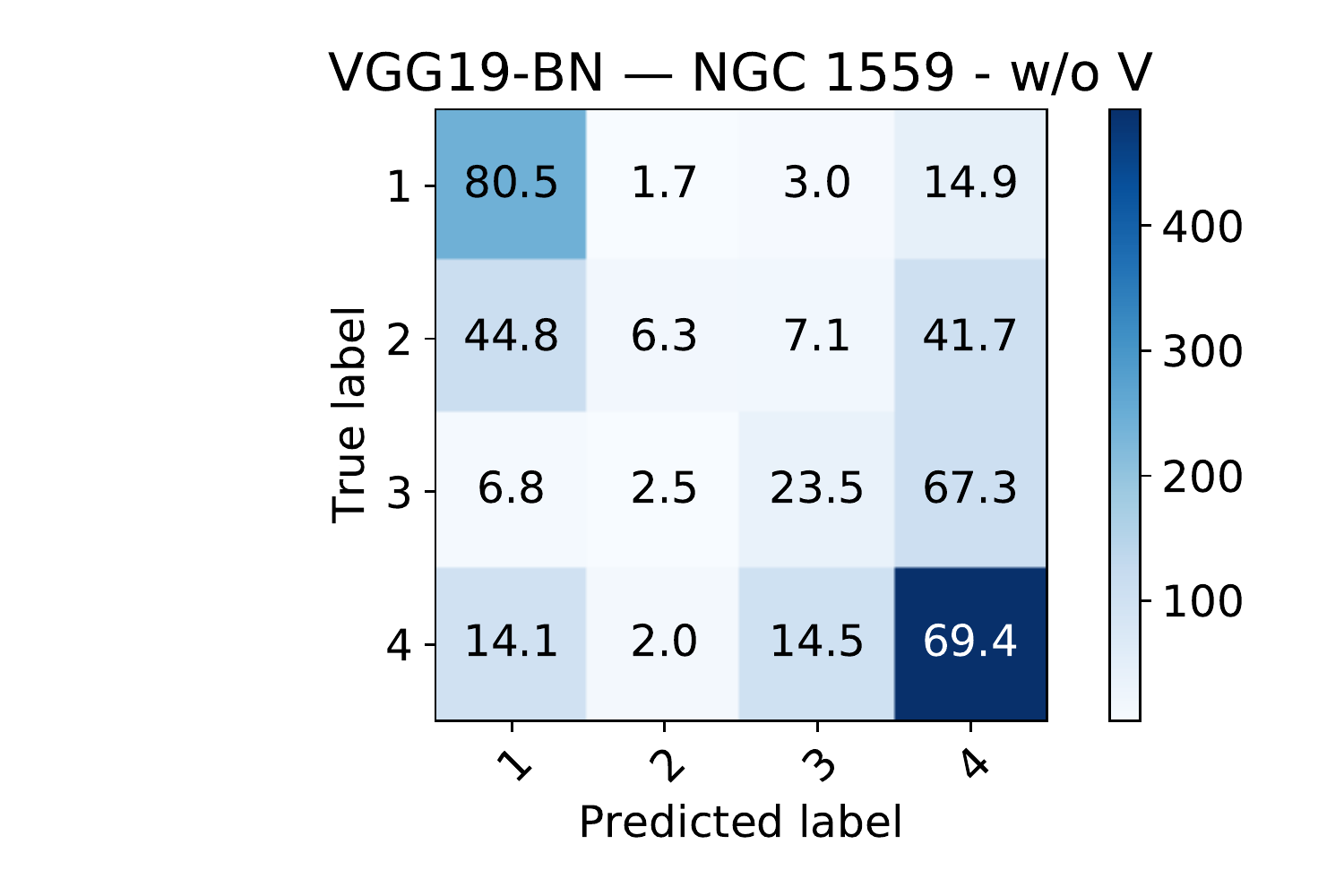}
 }
 \centerline{
    \includegraphics[width=.38\linewidth]{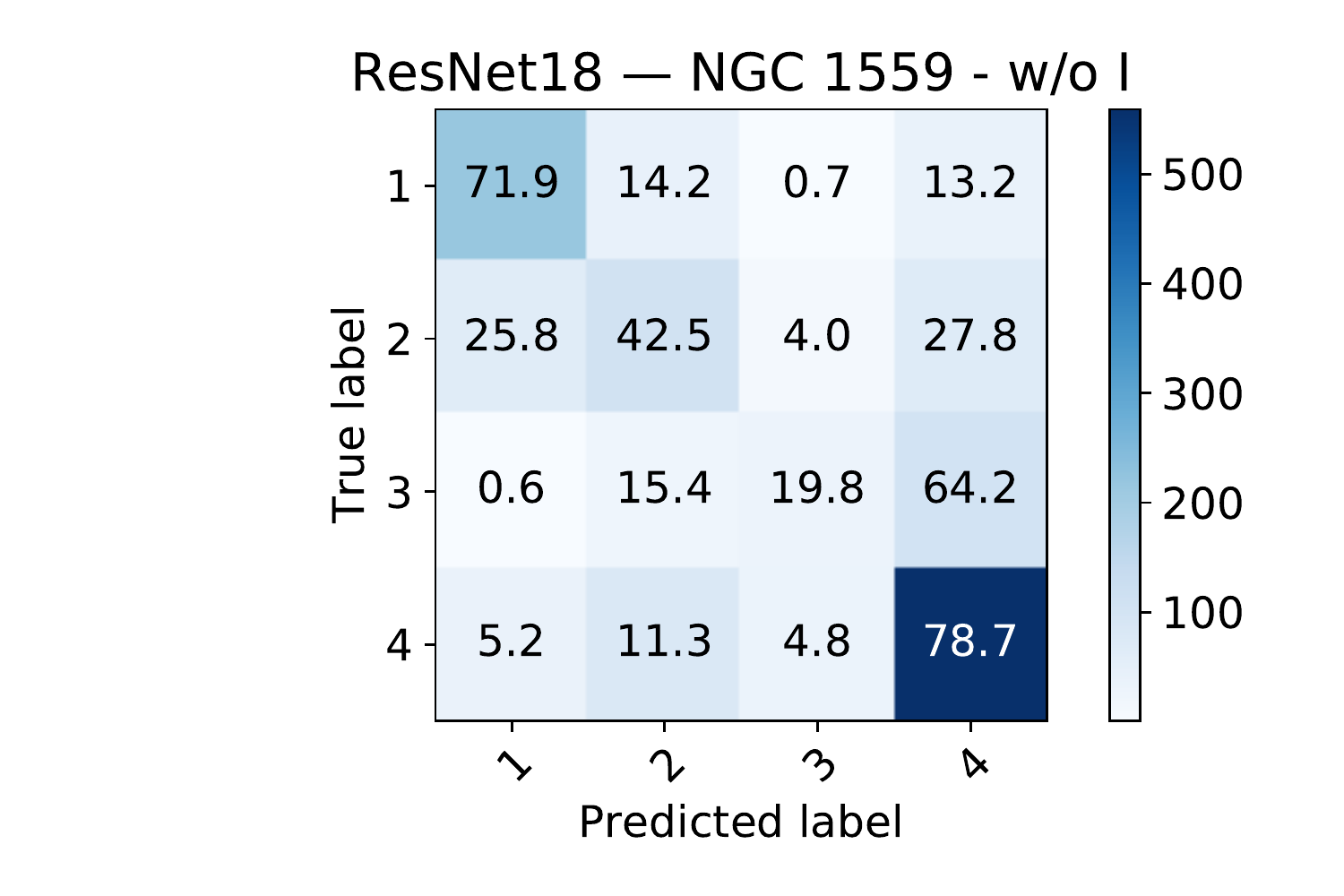}   
    \includegraphics[width=.38\linewidth]{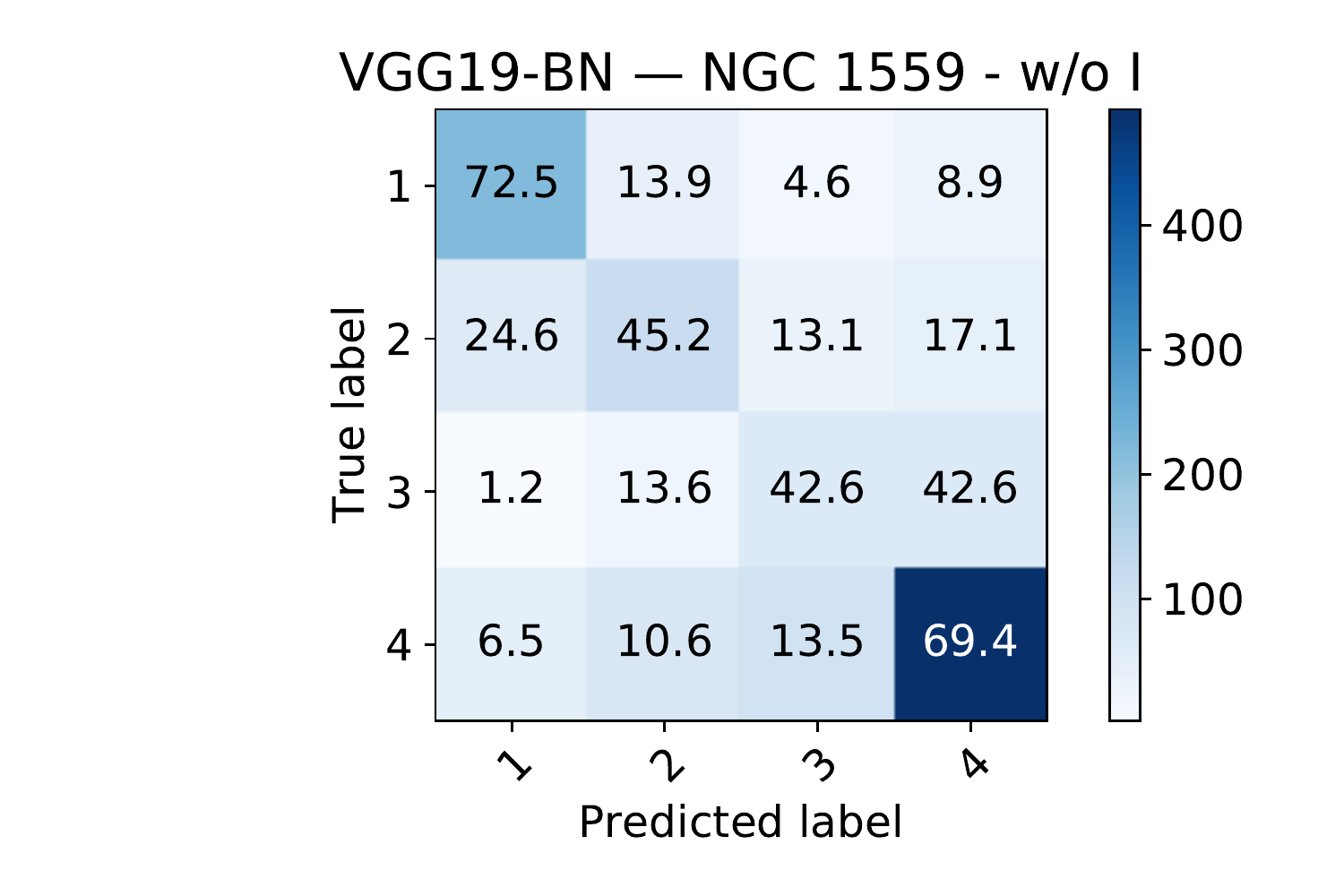}
 }
\caption{Left column: \texttt{ResNet} model classification results when the indicated filter is removed from the composite image. Right column: as before, but now for \texttt{VGG19-BN}. The greatest drop in the accuracies occurs when the V-band filter is removed.}
\label{fig:ablation_confusion}
\end{figure*}

We have also quantified what filter has the leading contribution for classification accuracy. To do so, we perform the following experiment: using NGC 1559 images as testing dataset, we produced five different testing datasets in which one filter was set to zero. We then fed these 5 different testing datasets, one at a time, to our neural network models trained with objects primarily classified by BCW and quantified which missing filter leads to the most significant drop in classification accuracy. As shown in Figure~\ref{fig:ablation_confusion}, the key filter is F555W. 

This finding is expected, since the human classifications primarily rely on the F555W image (e.g., using DS9 and imexamine), with color images (F814, F555, F336W) generated by the Hubble Legacy Archive providing supporting morphological information. Therefore, our neural network models seem to use insights similar to human vision to classify star cluster images.

\section{Discussion \& Conclusions}
\label{sec:end}
Using homogeneous datasets of human-labeled star cluster images from the Hubble Space Telescope, we have leveraged a new generation of neural network models and deep transfer learning techniques for morphological classification of compact star clusters in nearby galaxies to distances of $\sim$ 20 Mpc.  These results are very promising.

\begin{enumerate}
	\item Through all of the experiments presented here with multiple training sweeps for each neural network model, we see that the classification accuracy is similar for both architectures studied: i.e., ResNet18 and VGG19-BN pre-trained with the ImageNet dataset where the weights of the last layers and the last fully connected layers are randomly initialized. 
	\item Somewhat surprisingly, the performance of the models is relatively robust to the origin of the human classifications used, the particular galaxies included in the training sample, and the cropping size of the training images (spanning physical sizes of 16pc to 360pc).  Irrespective of whether the models are trained on a sample primarily classified by one expert (BCW) with galaxies at distances 2-4 times closer than the star cluster candidates to be evaluated in PHANGS-HST galaxy NGC 1559; or trained on the mode of classifications from three individuals where the sample does includes a galaxy at a distance similar to NGC 1559; the results are comparable.  The prediction accuracies for NGC 1559, which was not included in the training samples, are at the level of 70\%:40\%40-50\% for the class 1, 2, and 3 star clusters.  However, the BCW-trained networks have a higher performance in classification of the class 4 non-clusters in NGC 1559 (70\% vs. 50-60\%).  This might be expected since the classifications for NGC 1559 were also performed by BCW, and may be due to a higher level of self-consistency in the training and testing classification datasets. 
	\item Most importantly, despite training with relatively small datasets, the performance of the networks presented here is competitive with the consistency achieved in previous human and quantitative automated classification of the same star cluster candidate samples (Section~\ref{sec:c_human}).  Thus, this work provides a proof-of-concept demonstration that deep transfer learning can be successfully used to automate morphological classification of star cluster candidate samples 
	using HST UV-optical imaging being obtained by PHANGS-HST.
\end{enumerate}

\color{black}

This work represents a milestone in the use of deep transfer learning for this area of research, and represents progress from initial machine learning experiments described in \cite{grasha19} and also discussed in \cite{messa18}. \cite{grasha19} experimented with the use of an ML algorithm for classifying the approximately eleven thousand clusters in the spiral galaxy M51, based on a human classified training set with $\sim$2500 clusters from the LEGUS sample.  While the recovery of class 1 and 2 clusters is fairly good (in the range 60 - 75 \% in the Grasha and Messa studies, and comparable to the prediction accuracies presented here \color{black}) recovery of class 3 clusters is poor, with an apparently significant anti-correlation. 

To attempt to further improve upon the models presented here, future work will include training with the largest star cluster candidate sample possible (i.e., combining all samples used for this proof-of-concept demonstration plus classifications for objects in several galaxies in PHANGS-HST). \color{black} Improvement in classification accuracy also requires the development of a standarized dataset of human-labelled star cluster classifications, with classifications agreed upon by a full range of experts in the field, to be used as the basis for future network training.  This effort would benefit from a classification challenge, where experts can come to detailed agreement on the morphological features that constitute the criteria for classification (e.g., to establish full decision trees, such as those used for Galaxy Zoo by citizen scientists), \color{black} and explicitly describe where they disagree and why.  A review of differences in star cluster definitions between research groups, and their possible impact on conclusions about star cluster formation and evolution, can be found in \cite{krumholz18}.  The ultimate goal is to use deep learning techniques to not only rapidly produce reliable classifications and speed the time to science, but to significantly advance the field of star cluster evolution.  Given the discussion in \cite{krumholz18}, this requires that deep learning networks are trained on such standardized datasets, broadly adopted by workers in the field.  

With this study we open a new chapter to explore in earnest the use of deep transfer learning for the classification of very large datasets of star cluster galaxies in ongoing and future electromagnetic surveys, and application to the new PHANGS-HST data being obtained now.

\section*{Acknowledgements}

We thank the referee for feedback which significantly improved the paper, and in particular, motivated the expansion of our experiments to investigate potential differences in outcome when training with classifications by a single individual (BCW) versus using LEGUS consensus classifications from multiple individuals.  Initially, our experiments were based solely on the classifications of BCW.

We also thank the LEGUS team, and in particular Daniela Calzetti and Kathryn Grasha, for their pioneering efforts in the field of classifying star clusters using machine learning techniques, and for making results available via the LEGUS public website.  We thank Sean Linden for assisting BCW in classifications for star cluster candidates in NGC 3351, NGC 3627, and NGC 5457.  

Based on observations made with the NASA/ESA Hubble Space Telescope, obtained from the data archive at the Space Telescope Science Institute. STScI is operated by the Association of Universities for Research in Astronomy, Inc. under NASA contract NAS 5-26555.  Support for Program number 15654 was
provided through a grant from the STScI under NASA contract NAS5-
26555.

This research has made use of the NASA/IPAC Extragalactic Database (NED) which is operated by the Jet Propulsion Laboratory, California Institute of Technology, under contract with NASA. 

EAH and WW gratefully acknowledge National Science Foundation (NSF) awards OAC-1931561 and OAC-1934757.

This research is part of the Blue Waters sustained-petascale computing project, which is supported by NSF awards OCI-0725070 and ACI-1238993, and the State of Illinois. Blue Waters is a joint effort of the University of Illinois at Urbana-Champaign and its National Center for Supercomputing Applications. 

This work utilized resources supported by the NSF's Major Research Instrumentation program, grant OAC-1725729, as well as the University of Illinois at Urbana-Champaign.

We are grateful to NVIDIA for donating several Tesla P100 and V100 GPUs that we used for our analysis, and the NSF grants NSF-1550514, NSF-1659702 and TG-PHY160053. 

This research used resources of the Argonne Leadership Computing Facility, which is a DOE Office of Science User Facility supported under Contract DE-AC02-06CH11357. We thank the \href{http://gravity.ncsa.illinois.edu}{NCSA Gravity Group} for useful feedback. 

MC and JMDK gratefully acknowledge funding from the Deutsche Forschungsgemeinschaft (DFG) through an Emmy Noether Research Group (grant number KR4801/1-1) and the DFG Sachbeihilfe (grant number KR4801/2-1). JMDK gratefully acknowledges funding from the European Research Council (ERC) under the European Union's Horizon 2020 research and innovation programme via the ERC Starting Grant MUSTANG (grant agreement number 714907).

\bibliographystyle{mnras}   
\bibliography{all,ref_dl,dl_references}  


\appendix

\section{Statistical foundations of Deep Learning Classifiers}
\label{sec:stats}

Within the framework of statistical learning, an image $X$ can be modeled as a random matrix that takes value in set $\mathcal{X}$, and the corresponding class can be treated as a random variable $Y$ that takes value in set $\mathcal{Y}$. Since we use $299\times 299$ images with 5 channels, we treat a cluster image as random matrix of size $299\times 299\times 5$. Similarly, as we are trying to classify the images into 4 classes, $Y$ is a discrete random variable that takes values in $\mathcal{Y}$ with cardinality $|\mathcal{Y}|=4$.  

We assume that the star images and the corresponding class labels follow some unknown but fixed joint probability distribution, with the probability density function (pdf) $f_{XY}(x,y)$. We also use $\Delta_\mathcal{Y}$ to denote set of all possible distribution over $\mathcal{Y}$. Since in our case, $|\mathcal{Y}|=4$, we have $\Delta_\mathcal{Y}=\{\pi=(\pi_1,\pi_2,\pi_3,\pi_4):\sum_{i=1}^4\pi_i=1,\pi_i\geq 0, \forall i\in[4]\}$

Under these conventions, the goal of classification is to find a classifier or function $h: X\rightarrow \Delta_\mathcal{Y}$ that minimizes the expectation of the cross entropy between the predicted and the ground truth probability mass distribution (pmf) over the classes given the input image $X$, namely,

\begin{align}
    L(h) &=\mathbf{E}[H(h(X),f_{Y|X}(\cdot|X))]\\
    &= \int H(h(X),f_{Y|X}(\cdot|x)) f_{X}(x)dx\,,
\end{align}
where $f_{X}(x)$ is the marginal distribution of $X$ over $\mathcal{X}$, and $H$ is the cross entropy between the predicted and the ground truth pmf over classes,
\begin{align}
    H(h(x),f_{Y|Y}(\cdot|x)) = -\sum_{i=1}^4f_{Y|X}(Y=i|x)\log ([h(x)]_i),
\end{align}
and the $f_{Y|X}(y|x)$ is the conditional distribution of $Y$ given $X$.

In most cases, we only know the empirical distribution $\hat{f}_{XY}(x,y)$ of $(X, Y)$ and $\hat{f}_{Y|X}(y|x)$ of $Y$, which are determined by the empirical data. So the quantity we can directly minimize is

\begin{align}
    \hat{L}(h) &=\hat{\mathbf{E}}[H(h_X(\cdot),\hat{f}_{Y|X}(\cdot|X))]\\
    &= \int H(h_x(\cdot),\hat{f}_{Y|X}(\cdot|x)) \hat{f}_{X}(x)dx\,,
\end{align}

\noindent In practice, if the choice of $h(\cdot)$ is arbitrary, then finding an optimal solution is computationally unfeasible. Therefore, we often restrict the searching space to a class of parameterized functions, $h_{\mathbf{w}}(\cdot)$, where $\mathbf{w}$ is a vector of parameters. In this case, the optimization problem can be posed as 

\begin{align}\label{eq:min}
    \mathbf{w}^{*}=\argmin_{\mathbf{w}} \hat{L}(h_{\mathbf{w}})\,.
\end{align}

\noindent The choice of the parameterized function class is critical to the success of any statistical learning algorithm. In recent years, a deep-layered structure of functions has received much attention ~\citep{lecun2015deep,Goodfellow-et-al-2016},

\begin{align}{\label{eq:deepnet}}
    h_{\mathbf{w}}(\mathbf{x})=h_{\mathbf{w}_n}(h_{\mathbf{w}_{n-1}}(\cdots h_{\mathbf{w}_1}(\mathbf{x}))),
\end{align}

\noindent where \(n\) is the number of layers or the depth. Usually, we choose, \(h_{\mathbf{w}_i}(\mathbf{x})=g(\mathbf{w}_i\mathbf{x})\), where \(\mathbf{w}_i\) is a matrix, \(\mathbf{x}\) is an input vector, and \(g(\cdot)\) is a fixed non-linear function, e.g., $\max\{\cdot,0\}$ (also known as \texttt{ReLU}), \(\tanh(\cdot)\), etc, that is applied element-wise. For the classification problems, we usually apply the so-called softmax function after the last linear transformation. The softmax function on a vector $\mathbf{x}$ is a normalization after an element-wise exponentiation,
\begin{align}\label{eq:softmax}
    \text{softmax}(\mathbf{x})_i=\frac{\exp(x_i)}{\sum_{i=1}^n{\exp(x_i)}},\ \ \ \ \forall i=1,...,n,
\end{align}
where $n$ is the length of $\mathbf{x}$.

This function class and its extensions, also dubbed neural networks, combined with simple first-order optimization algorithms such as stochastic gradient descent (SGD), and improved computing hardware, has lead to disruptive applications of deep learning~\citep{lecun2015deep,Goodfellow-et-al-2016}.

\section{Deep transfer learning}
\label{deeptransfer}
In practice, Eq.~\ref{eq:min} is usually iteratively solved by using variants of SGD. Thus, the choice of initial value for weights $\mathbf{w}$ is critical to the success of the training algorithm. If we have some prior knowledge about what initial wights $\mathbf{w}_0$ works better, then it is highly possible that the numerical iteration can converge faster and return better weights $\mathbf{w}$. This is the idea behind deep transfer learning~\citep{bengio2011deep,Goodfellow-et-al-2016}.

For a deep learning neural network, such as the one defined by Eq.~\ref{eq:deepnet}, the layered structure can be intuitively interpreted as different levels of abstraction for the learned features. In other words, layers that are close to the input learn lower-level features, such as different shapes and curves in the image, and layers that are close to the final output layer learn higher-level features, such as the type of the input image.  Suppose we have a trained model that works well in one setting, with probability distribution $f^{(1)}_{XY}$, and now we would like to train another model in a different setting, with with probability distribution $f^{(2)}_{XY}$. If the images drawn from the distributions $f^{(1)}_{XY}$ and $f^{(2)}_{XY}$ share some features, then it is possible to transfer weights from the model trained on images sampled from $f^{(1)}_{XY}$, to the model that we would like to train, using images sampled from $f^{(2)}_{XY}$, with the assumption that the weights from the  model trained on images sampled from $f^{(1)}_{XY}$, can also be useful in extracting features from images drawn from the distribution $f^{(2)}_{XY}$. So, instead of training the second model from scratch, we can initialize the weights of the second model to those of the first model that we trained in a different setting (e.g., distribution $f^{(1)}_{XY}$), and utilize the common features we have already learned in the previous setting.

\section{Batch normalization}
\label{batchnorm}
The weights of each layer in a neural network model change throughout the training phase, which implies that the activations of each layer will also change. Given that the activations of any given layer are the inputs to the subsequent layer, this means that the input distribution changes at every step. This is far from ideal because it forces each intermediate layer to continuously adapt to its changing inputs.Batch normalization is used to ameliorate this problem by normalizing the activations of each layer.In practice this is accomplished by adding two trainable parameters to each layer, so the normalized output is multiplied by a standard deviation parameter, and then shifted by a mean parameter. With this approach only two parameters are changed for each activation, as opposed to losing the stability of the network by changing all the weights. It is expected that through this method each layer will learn on a more stable distribution of inputs, which may accelerate the training stage.

\bsp	
\label{lastpage}
\end{document}